%% file: ToN_AttackFINAL.tex
\documentclass[journal,comsoc]{IEEEtran}
\usepackage[T1]{fontenc}

\ifCLASSINFOpdf
\else
\fi
\usepackage{amsmath,bm}
\usepackage[cmintegrals]{newtxmath}
\usepackage{url}

\usepackage{bm}
\usepackage{graphicx,amsmath}
\usepackage{subfigure,epsfig}
\usepackage{epstopdf}
\usepackage{color}
\usepackage{amssymb}
\usepackage{latexsym}
\usepackage{algorithm}
\usepackage{algpseudocode}

\input{./pream}
\input{./math_def}

\graphicspath{{Figures/}}

\begin{document}

\title{Attack vulnerability of power systems under an equal load redistribution model}


\author{Talha Cihad Gulcu, \and Vaggos Chatziafratis, Yingrui Zhang, and Osman~Ya\u{g}an
       \thanks{This research was supported by the National Science Foundation through grant CCF \#1422165 and a generous gift from {\em Persistent Systems, Inc}.} 
            \thanks{Preliminary version of some of the material was presented 
at the Information Theory and Applications Workshop (ITA 2016), San Diego, CA 
\cite{vaggos_ITA}.}
       \thanks{T. C. Gulcu is with TUBITAK Software Technologies Research Institute, 06100,
Cukurambar, Ankara, Turkey. E-mail: tcgulcu@gmail.com}
       \thanks{V. Chatziafratis is with the
       Computer Science Department,
Stanford University, Palo Alto, CA 94305.
Email:vaggos@stanford.edu}
\thanks{Y. Zhang and O. Ya\u{g}an are with the Department
of Electrical and Computer Engineering,
Carnegie Mellon University,
Pittsburgh, PA 15213. E-mail:\{yingruiz, oyagan\}@andrew.cmu.edu}
}


\markboth{IEEE/ACM Transactions on Networking,~Vol.~X, No.~X, 2018}%
{Shell \MakeLowercase{\textit{et al.}}: Bare Demo of IEEEtran.cls for IEEE Communications Society Journals}

\maketitle

\begin{abstract}
This paper studies the vulnerability of flow networks against adversarial attacks. 
In particular, consider
a power system (or, any system carrying a physical flow) consisting of $N$ transmission lines with initial loads $L_1, \ldots , L_N$ and capacities $C_1, \ldots, C_N$, respectively; the capacity $C_i$
defines the maximum flow allowed on line $i$. Under an equal load redistribution model, where load of failed lines is redistributed equally among all remaining lines, we study the   {\em optimization} problem of finding the best $k$ lines to attack so as to minimize the number of {\em alive} lines at the steady-state (i.e., when cascades stop). This is done to reveal the worst-case attack vulnerability of the system as well as to reveal its most vulnerable lines. We derive optimal attack strategies in several special cases of load-capacity distributions that are practically relevant. 
We then consider a modified optimization problem where the adversary is also constrained by the {\em total} load (in addition to the number) of the initial attack set, and prove that this  problem is NP-Hard. 
Finally, we develop  heuristic algorithms for selecting the attack set for both the original and modified problems.  Through extensive simulations, we show that these heuristics outperform benchmark algorithms under a wide range of settings.
 \end{abstract}
\vspace{-2mm}

\vspace{-1mm}

\section{Introduction}

Networks that carry or transport physical commodities, e.g.,  electricity/water/gas distribution networks and road/public transportation networks, have been an integral part of our daily lives for decades. For example,
our quality of life largely depends on the continuous availability
of an electrical power supply. This dependence is expected to be further amplified in the near future due to 
the increasing market share of electric vehicles and increasing integration of major national infrastructures to the 
power grid; e.g., water, transport, communications, etc.
All of these point to a future where the reliability of the flow networks will be paramount, with the central research question being how we can design  these networks in a robust and reliable manner. In the rest of the paper, we focus our attention  on power systems for concreteness, but  most of the discussion applies generally to any transport system.

A major problem with existing power systems is the seemingly unexpected large scale failures. 
Although rare, the sheer size of such failures has proven to be very costly, 
at times affecting hundreds of millions of people \cite{andersson2005causes,rosas2011analysis,yang2017small}; e.g., the recent blackout in India \cite{zhang2013understanding}. 
Such events are often attributed to a 
small initial shock getting escalated due to intricate dependencies within a power system
\cite{Buldyrev,WattsExternal,kinney2005modeling}. 
This phenomenon, also known as cascade of failures, 
has the potential of collapsing an entire power system as well as
other infrastructures that depend on the power grid  \cite{dobson2007complex,o2007critical,yagan_fiber_bundle}. 
Thus, understanding the dynamics of failures in power systems 
and mitigating the potential risks are critical for the successful
development and evolution of many critical infrastructures.

In this paper, we continue our study of the robustness of power systems under a simple model based on 
 {\em equal} redistribution of flow upon the failure of a power line. 
Namely, we consider a power system with $N$ transmission lines
with initial loads $L_1,\ldots, L_N$ and capacities $C_1,\ldots, C_N$. 
If a line fails (for any reason), its load is assumed to be redistributed equally among all lines that are {\em alive}.
Thus, the load carried by a line $i$ may exceed its initial value $L_i$ over time due to load redistribution. 
The capacity $C_i$ defines the maximum flow allowed on the line $i$, meaning that if 
the load carried by $i$ exceeds this capacity at any time, the line will be tripped (i.e., disconnected) by means of automatic protective equipments so as to avoid costly damages to the system. Subsequently, the load that was carried by line $i$ before failure will be redistributed to remaining lines, which in turn may cause further failures, possibly leading to a {\em cascade} of failures. 


The equal load redistribution model gets its appeal from its ability to capture the {\em long-range} \cite{daqing2014spatial,PahwaScoglioScala} nature of the Kirchhoff's law, at least in the mean-field sense, as opposed to the {\em topological} models \cite{CrucittiLatora,MotterLai,WangChen} where failed load is  redistributed only {\em locally} among neighboring lines.
For example, it was  suggested by Pahwa et al. \cite{PahwaScoglioScala} that equal load redistribution is a reasonable
assumption\footnote{The equal load redistribution model is also similar in spirit to the well-studied CASCADE model introduced in \cite{dobson2007complex,dobson2002examining}.
There, 
upon failure of a line, a fixed amount $\Delta$ of load is redistributed to all functional lines irrespective of the load being carried by the failed line or the number of remaining lines.} 
especially under  
the DC power flow model;  the DC model is  known 
\cite{Overbye,StottJardimAlsac} to approximate the AC model well in many cases. 
With these in mind, an important goal is to understand the robustness of systems under the equal load redistribution model described
above against {\em random} and {\em targeted} attacks. 
The former case was studied by Ya\u{g}an \cite{yagan_fiber_bundle} 
under the assumptions that initial loads $L_1,\ldots, L_N$ are independent and identically distributed with 
$P_L(x) = \bP{L \leq x}$ and that capacities are given by
\[ 
C_i = (1+\alpha) L_i, \quad i=1,\ldots, N,
\vspace{-1mm}
\]
where $\alpha>0$ denotes the {\em tolerance factor}; in \cite{yagan_fiber_bundle} 
all lines assumed to have the same tolerance factor. There,
the robustness of the system against {\em random} attacks that target a $p$-fraction of the lines was studied;
system robustness was quantified by the  {\em final} (i.e., steady-state) 
fraction $n_{\infty}(p)$ of {\em non-failed} lines.
Among other results, it was shown that the system robustness
is maximized  
if all lines are given the same initial load, for a given fixed mean load $\bE{L}$. 

Recently, Zhang and Ya\u{g}an  \cite{yingrui_yagan_optimal} extended the results in \cite{yagan_fiber_bundle}
to the more general case where lines can have varying tolerance parameters. Namely, they let
\vspace{-1mm}\[
C_i =  L_i + S_i, \quad i=1,\ldots, N,
\vspace{-1mm}
\]
with $S_i$ denoting the free-space (or, redundancy) available at line $i$. The tolerance factor,
given by $\alpha_i = S_i/ L_i$, may now vary from one line to another.  Under the assumption that load-\lq free space' pairs
$(L_i, S_i)$ are independent and identically distributed with 
$P_{LS}(x,y) = \bP{L \leq x, S \leq y}$, they studied 
 the robustness against random attacks that target a $p$-fraction of the lines.
Their main result is that, with the mean values $\bE{L}$ and $\bE{S}$ are fixed, 
$n_{\infty}(p)$ is uniformly maximized for all $p$ values if all nodes are given the same free space $\bE{S}$, 
regardless of how the loads are distributed. 
This leads to the counterintuitive conclusion that lines with higher initial load shall be assigned smaller tolerance factors to maximize robustness  and raises the possibility that the widely used (both in academy and industry) setting of equal tolerance factor is not optimal for system robustness (as far as the metric $n_{\infty}(p)$ is concerned).

With the case of random attacks being  well-understood, we shift our attention in this paper to understanding the vulnerability of power systems under {\em targeted} attacks.
As before, the main goal would be to derive design strategies (in the form of optimal load-\lq free space' distributions) 
that would lead to maximum robustness, this time against a knowledgeable adversary attacking a carefully selected set of lines.
However, for this optimization problem to be well-defined one has to have a good understanding of the problem from an adversary's perspective. 
With this in mind, this paper aims to develop {\em good} attack strategies that lead to maximal damage to the system for a given number of lines that can be attacked. The solution to the optimal attack problem will also help a system designer by i) revealing the worst-case attack vulnerability of the system
which can help evaluate a given system design; and ii) revealing the most vulnerable lines in the system that will potentially be targeted by adversaries; this may then provide useful design guidelines for {\em improving} system robustness. 

Formally, we consider the following optimization problem. 
Given $N$ lines with loads $L_1, \ldots, L_N$ and free spaces $S_1,\ldots, S_N$, we  seek to find the optimal set $A$ of $k$ lines that the adversary should attack 
in order to minimize the final number $n_{\infty}(A)$ of alive lines. We provide optimal solutions via greedy algorithms in 
three special cases: i) when all lines have the same load; ii) when $S_i = \alpha L_i$ for each $i=1,\ldots, N$ (as commonly used in the literature \cite{CrucittiLatora,kinney2005modeling,Mirzasoleiman,MotterLai,WangChen,yagan_fiber_bundle});
and iii) when all lines have the same free space, i.e., when $S_1 = \cdots = S_N$.  The last case is of particular interest as it is known to lead to maximum robustness against random attacks. 
For general cases of load-\lq free-space' distribution $P_{LS}$, we develop heuristic attack algorithms and show via simulations that these outperform benchmark attack strategies. In particular, we show that attacking the lines with the {\em largest} Load $\times$ Free-space ($L*S$) is in general a good strategy. Parametrized attack strategies targeting the lines with the largest $L*S^{\beta}$ are also considered where $\beta$ is in $[0, \infty]$, and optimum choices for $\beta$ are provided in several cases.

We also consider a variation of the problem
with an additional constraint on the total load of the lines attacked; i.e., when the adversary is further constrained with $\sum_{i \in A} L_i \leq Q$ for some $Q$. From a practical point of view, this might be the case if high-load carrying lines are {\em protected} better by the network owner and the {\em cost} of attacking them is proportional to their load.  
We show that this variation of the optimal attack problem
is in fact NP-Hard, meaning that no polynomial-time algorithm can find the  set $A$ that minimizes $n_{\infty}(A)$, unless $P \equiv NP$.
Our proof is based on a polynomial time reduction from the {\em $k$-Subset Sum} problem, i.e., the problem that seeks to find whether a sequence of
integers has a subset of size $k$ whose sum equals $Q$. For the modified optimization problem, we develop several heuristic algorithms and evaluate their performance in comparison with benchmarks through an extensive simulation study. In particular, we modify the previously developed heuristics with a {\em switch} that, when actuated during a sequential selection of lines to be attacked, changes the way algorithm makes the remaining selections; this idea is inspired from heuristics developed in \cite{loulou1979new} for the multi-dimensional $0$-$1$ Knapsack problem. Among other things, we demonstrate via  simulations that the max-$L*S$ algorithm with a {\em switch} performs well in a range of settings. 

We believe that our results will shed some light on the vulnerability of power systems against powerful adversaries launching {\em targeted} attacks. In particular, we expect them to capture the qualitative properties of a power system well. 
This work may also have applications in other fields. In particular, the model considered here might be relevant for any flow network that is responsible for carrying (or, transporting) a physical quantity.  A particularly interesting application is the study of traffic jams in roads \cite{pradhan2003failure}, where the capacity of a line is regarded as the traffic flow capacity of a road.


The rest of the paper is organized as follows. We describe the system model in Section
\ref{sec:Model}. In Section \ref{sec:survey}, we survey the results in \cite{yingrui_yagan_optimal} concerning robustness against random attacks.
We start our discussion on optimal attack strategies in Section \ref{sec:attack_perspective}
by demonstrating that certain greedy algorithms fail to give the optimal solution in general.  
In Section \ref{sec:greedy_success} we consider some special cases of interest where greedy algorithms are shown to 
find optimal attack sets. In Section \ref{sec:hardness} we prove a hardness result showing that a variation of the
optimal attack problem is NP-Hard. Finally, in Section \ref{heuristic}, we develop heuristic algorithms for both the original and the modified optimization problems and present a detailed numerical study comparing the performance of our heuristics with benchmark algorithms such as max-$S$, max-$L$, max-$C$, and random attacks.

The random variables (rvs)
under consideration are defined on the probability space
$(\Omega, {\cal F}, \mathbb{P})$, where
 $\mathbb{P}$ denotes the probability measure. We
denote the  expectation operator by $\mathbb{E}$.
The indicator function of an event $E$ is denoted by $\1{E}$.  We write $|A|$ for the cardinality
of a discrete set 
$A$.


\section{System Model}
\label{sec:Model}
We consider a power system with $N$ transmission lines $\mathcal{L}_1, \ldots, \mathcal{L}_N$ 
with initial loads (i.e., power flows) $L_1, \ldots, L_N$.
The {\em capacity} $C_i$  of a line $\mathcal{L}_i$ defines the maximum power flow that it can sustain, and 
is  given by 
\begin{equation}\vspace{-1mm}
C_i= L_i + S_i, \qquad i=1,\ldots, N,
\label{eq:capacity}
\end{equation}
where $S_i$ denotes the {\em free-space} (or, redundancy) available to line $\mathcal{L}_i$. 
The capacity of a line is typically written \cite{MotterLai,WangChen,Mirzasoleiman,CrucittiLatora}   as a  factor of the line's original load, i.e.,
\[\vspace{-1mm}
C_i = (1+\alpha_i) L_i
\]
with $\alpha_i>0$ defining the {\em tolerance} parameter for line $\mathcal{L}_i$. Put differently, the free space $S_i$ is often given in terms of
the initial load $L_i$ with $S_i = \alpha_i L_i$. 
It is assumed that a line {\em fails} (i.e., outages) if its load exceeds its capacity at any given time. In that case, the load it was carrying before the failure
is redistributed {\em equally} among all remaining lines. 

The main question of interest is to characterize the robustness of this power system against i) {\em random}  attacks that result initially with a failure of a (randomly selected) $p$-fraction of the lines; and ii) {\em targeted} attacks
that initially fail a specific set $A$ of lines.
The initial set of failures lead to redistribution of power flows from the failed lines to {\em alive} ones (i.e., non-failed lines), 
so that the load on each alive line becomes equal to its initial load plus its equal share of the total load of the failed lines.  This may lead to 
the failure of some additional lines due to 
the updated flow exceeding their capacity. This process may continue recursively, generating a {\em cascade of failures}, 
with each failure further increasing the load on the alive lines, and may eventually result
with the collapse of the entire system. 

Throughout, let $n_{\infty}(p)$ denote the {\em final} fraction of alive lines
when a $p$-fraction of lines is randomly attacked. The robustness of the power system can be evaluated \cite{yagan_fiber_bundle,yingrui_yagan_optimal}
by the behavior of $n_{\infty}(p)$ as the attack size $p$ increases, and particularly by the critical attack size $p^{\star}$ at which 
$n_{\infty}(p)$ drops to zero. In the case where a specific set $A$ of lines are attacked, we define $n_{\infty}(A)$ as the final {\em number} of alive lines. The main goal of this paper is to seek {\em optimal} attack strategies, i.e., to find the set $A$ of lines that minimizes $n_{\infty}(A)$ under certain constraints; e.g., the size $|A|$ being fixed. 

The equal flow redistribution model is inspired by the democratic fiber bundle model \cite{AndersenSornetteKwan,Daniels1945},
where $N$ parallel fibers with random failure thresholds $C_1, \ldots, C_N$ (i.e., capacities) drawn independently from $P_C(x)$ share equally an applied total force of $F$; see also 
\cite{pradhan2003failure,roy2015fiber}. This model has been recently adopted by Pahwa et al. \cite{PahwaScoglioScala} in the context 
of power systems
with $F$ corresponding to the total load that $N$ power lines share equally.
The  formulation considered here, introduced in \cite{yagan_fiber_bundle,yingrui_yagan_optimal},
differs from the original democratic fiber-bundle model in that
i) we do not assume that the total load of 
 the system is fixed at $F$; and ii) we allow for power lines to carry different initial loads;
 see also \cite{ozel2018robustness}.
 In addition,  \cite{PahwaScoglioScala} is concerned with failures in the power system 
 that are triggered by increasing the total force (i.e., load) applied.
Instead, our formulation allows analyzing the robustness of the system against external attacks
or random line failures, which are known to be the source of system-wide blackouts in many interdependent systems \cite{Buldyrev,YaganQianZhangCochranLong,ZhangArenasYagan}.

\section{Relevant Work: Evaluating and Optimizing Robustness against Random Attacks}
\label{sec:survey}

We now survey the results obtained by  Zhang and Ya\u{g}an \cite{yingrui_yagan_optimal,yagan_fiber_bundle}
 on the robustness of power systems under equal load redistribution of loads. These works consider the problem from a defender's perspective and provide means to characterize and optimize the robustness of the system, assuming that the adversary will launch a {\em random} attack to a certain fraction of lines. With the randomness involved in the attack model, as well as load-capacity values,   \cite{yagan_fiber_bundle, yingrui_yagan_optimal} rely on {\em mean-field} analysis to characterize the {\em mean} performance of the system in the asymptotic regime where $N$ approaches infinity.
 
 Throughout this section, assume that the pairs  $(L_i, S_i)$ are independently and identically distributed with $P_{LS}(x,y):=\bP{L \leq x, S \leq y}$ for each  $i=1,\ldots, N$. The corresponding probability density
 is given by $p_{LS}(x,y) = \frac{\partial^2}{\partial x \partial y} P_{LS}(x,y)$.
We assume that  $L \geq L_{{\sl min}} > 0$ and $S>S_{{\sl min}}>0$ with probability one, and 
 that the marginal densities $p_L(x)$ and $p_S(y)$ are continuous on their support.

\subsection{Final system size as a function of the attack size}
The main result in \cite{yingrui_yagan_optimal}
characterizes the robustness of power systems under any distribution of initial load $L$ and free space $S$, and
any attack size $p$. 
\begin{thm}[\cite{yingrui_yagan_optimal}]
{\sl Let $L$ and $S$ denote generic random variables following the same distribution with initial loads $L_1,\ldots,L_N$,
and free space $S_1, \ldots, S_N$, respectively.
Then, with $x^{\star}$ denoting the smallest solution of
\begin{equation}
\bP{S > x} \left( x + \bE{L ~|~ S > x} \right) \geq \frac{\bE{L}}{1-p}
\label{eq:main_condition}
\end{equation}
over $x \in (0, \infty]$, the final system size $n_{\infty}(p)$ is given by 
\begin{equation}
n_{\infty}(p) = (1-p) \bP{S > x^{\star}}.
\label{eq:final_size}\vspace{-4mm}
\end{equation}
 }
 \label{thm:final_size}
\end{thm}

 An intuitive explanation of this result is that $x^{\star}$, defined as the smallest solution of (\ref{eq:main_condition}), gives the {\em extra load} each alive line has to carry (in addition to their initial load) at the steady-state;  (\ref{eq:final_size}) is then understood easily since a line will still be alive at the steady-state if i) it is {\em not} targeted initially; and ii) it has enough free-space to handle the extra $x^{\star}$ amount of load. 

\begin{figure}[!t]
\vspace{-3mm}
\hspace{-1mm}
\includegraphics[width=0.48\textwidth] {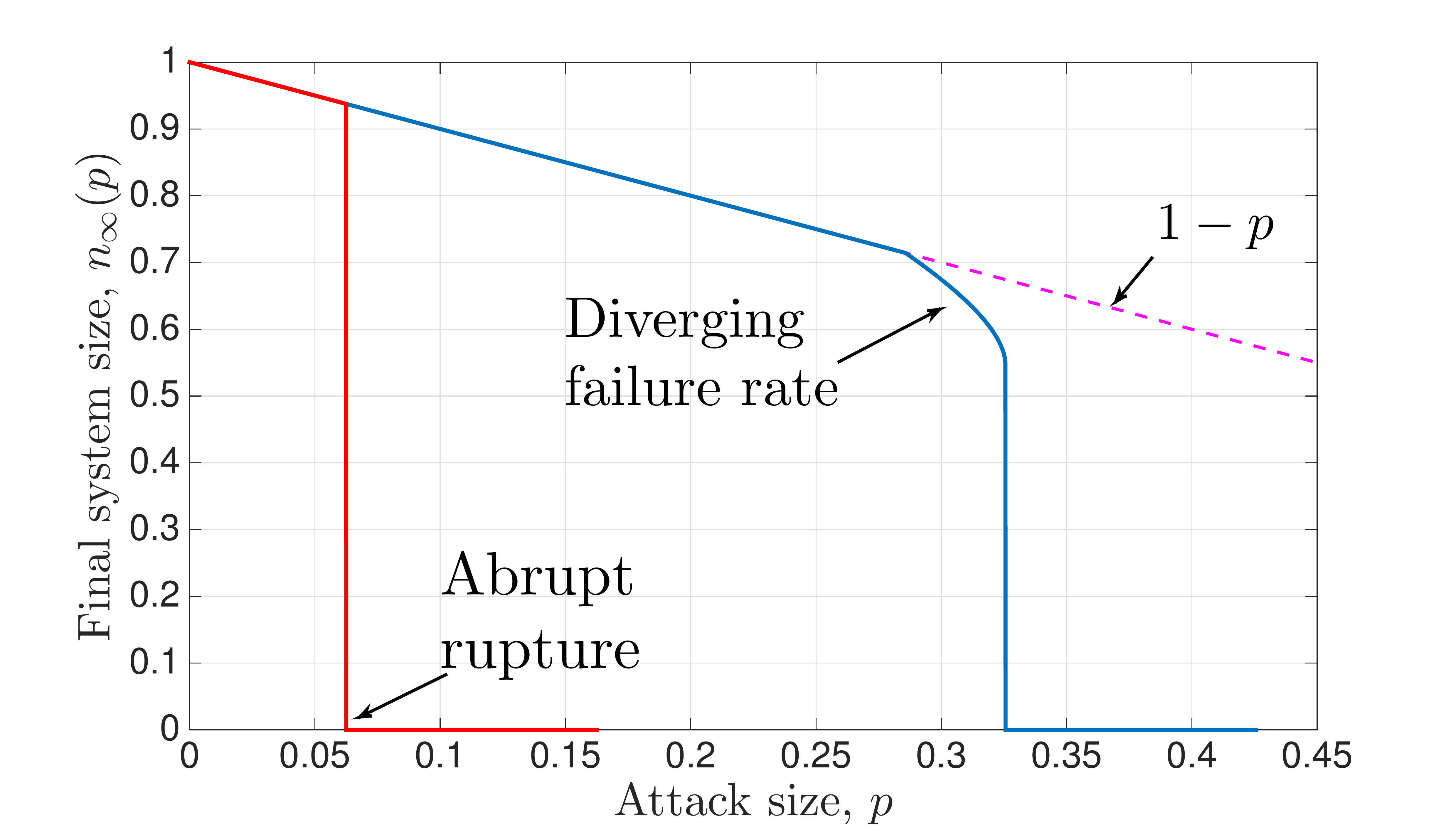}
\vspace{-2mm}
 \caption{ \sl We see the variation of final system size $n_{\infty}(p)$ as a function of the (random) attack size $p$
  when
  loads are drawn independently and uniformly from the range
 $[L_{\textrm{min}},L_{\textrm{max}}] = [10,50]$, and free spaces  are given by $S_i = \alpha L_i$.
Red (lower) curve stands for  $\alpha = 0.2$, whereas 
 blue (upper) curve represents $\alpha=1.2$.  
For $\alpha=0.2$, we observe an abrupt first-order transition of $n_{\infty}(p)$ as it suddenly jumps to zero at  $p^{\star} = 0.0625$, while decaying linearly as $1-p$ up until that point.
A slightly different behavior is seen when  $\alpha=1.2$ in that $n_{\infty}(p)$ exhibits a continuous divergence from the $1-p$ line before complete system failure, which again takes place through a discontinuous transition. 
  }\vspace{-3mm}
\label{fig:examples}
\end{figure}

For a graphical solution of $n_{\infty}(p)$, 
one shall plot $\bP{S > x} \left( x + \bE{L ~|~ S > x} \right)$ as a function of $x$, and draw a horizontal line at $\bE{L}/(1-p)$ on the same plot. The leftmost intersection of these two lines gives the operating point $x^{\star}$, from which we can compute  $n_{\infty}(p) = (1-p)\bP{S > x^{\star}}$. When there is no intersection,  we set $x^{\star}=\infty$ and understand that $n_{\infty}(p)=0$; e.g., see Figure \ref{fig:examples}.

Theorem \ref{thm:final_size} enables computing $n_{\infty} (p)$ as a function of $p$, thereby helping understand the response of the system to attacks of varying magnitude. It also enables computing the critical attack size $p^{\star},$ defined as the largest attack the system can sustain. More precisely, we define
\[
p^{\star} = \sup_{p>0} n_{\infty}(p) > 0, 
\]
so
that for any attack with size $p > p^{\star}$, the system undergoes a complete breakdown leading to 
$n_{\infty} (p) = 0$. The next result provides a closed form expression for the critical attack size. 
\begin{thm}[\cite{yingrui_yagan_optimal}]
{\sl The critical attack size $p^{\star}$ is 
given by
\begin{equation}
p^{\star} = 1- \frac{\bE{L}}{\max\limits_{x}\{\bP{S > x} \left( x + \bE{L ~|~ S > x} \right)\}}.
\label{eq:max_attack}
\end{equation} 
}
\label{thm:max_attack}
\vspace{-1mm}
\end{thm}
The operational usefulness of $p^{\star}$ derives in part from its ability to quantify the system robustness by a single scalar, paving the way to designing the system with optimal robustness by means of maximizing $p^{\star}$; see  Section \ref{subsec:optimal_robustness} below.

An interesting
question here is whether $n_{\infty}(p)$ decays to zero continuously (i.e., through a second-order transition),
or discontinuously (i.e., through a first-order transition). The practical significance of this is that continuous transitions  suggest a more stable and predictable system behavior with respect to attacks, whereas with discontinuous transitions, system behavior becomes more difficult to predict based on past data.
In \cite{yingrui_yagan_optimal}, it was shown 
 that the total breakdown of the system will always be through a {\em first-order} (i.e., discontinuous) transition under random attacks. 
More precisely, it was shown that 
$
n_{\infty} (p^{\star}) > 0,
$
while by definition it holds that
$
n_{\infty} (p^{\star}+\varepsilon) = 0
$ 
for any $\varepsilon>0$ arbitrarily small.
This shows that despite its simplicity, the equal flow redistribution model can capture  real-world phenomena 
of unexpected large-scale system collapses; i.e., cases where seemingly identical attacks/failures leading to entirely different consequences. In addition, depending on system parameters, the first-order breakdown may have early indicators at smaller attack sizes such as a 
{\em diverging} failure rate leading to a non-linear decrease in  $n_{\infty} (p)$. In other cases, the system {\em abruptly} transitions to a total collapse while perfectly resisting  attacks until the critical point; see Figure \ref{fig:examples} and \cite[Fig.~2]{yingrui_yagan_optimal} for 
 the rich transition behavior that can be observed under the equal load redistribution model.

%
%
%
%

\subsection{Achieving {\em Optimal} Robustness}
\label{subsec:optimal_robustness}
An important question from a system designer's perspective is 
concerned with deriving the {\em universally optimum} distribution of initial loads $L_1, \ldots L_N$ 
and free spaces $S_1,\ldots, S_N$
when the mean values $\bE{L}$ and $\bE{S}$, respectively, are fixed. 
The  results obtained in  \cite{yingrui_yagan_optimal} concerning this   are presented next.
\begin{thm}[\cite{yingrui_yagan_optimal}]
{\sl 
For any initial load-free space distribution, it holds that
\begin{equation}
p^{\star} \leq \frac{\bE{S}}{\bE{S}+\bE{L}} = \frac{\bE{S}}{\bE{C}} = p^{\star}_{\textrm{dirac}} 
\label{eq:max_attack_upper}
\end{equation}
where $p^{\star}_{\textrm{dirac}}$ denotes the critical attack size
when the free-space follows a {\em Dirac}-delta distribution, i.e.,
$P_{LS}(x,y) = P_{L}(x) \1{y \leq \bE{S}}$, for any $P_{L}(x)$.
}
\end{thm}

In words, this result states that the critical attack size can never be larger than the ratio of mean free space to
mean capacity. 
In addition, this {\em optimal} $p^{\star}$ value can be achieved by 
assigning every line the same free-space, i.e., by setting
 $S_1= \cdots = S_N = \bE{S}$, regardless
 of how loads are distributed.


Thus far, we have seen that the equal distribution of free space leads to the largest possible critical attack size,
hereafter denoted $p^{\star}_{\textrm{opt}}$. It is clear that the final system size after an attack of size $p$ is at most $1-p$. With this
in mind, the next result establishes the optimality of the Dirac-delta distribution of free-space in the sense of maximizing the robustness
of power systems {\em uniformly} over all attack sizes. 
\begin{thm}[\cite{yingrui_yagan_optimal}]
{\sl With $S_1=S_2= \cdots = S_N$ we have
\[
n_{\infty}(p) = \left \{  
\begin{array}{cc}
1-p  &  \textrm{for $p < p^{\star}_{\textrm{opt}} $}   \\
0  &     \textrm{for $p \geq p^{\star}_{\textrm{opt}} $}
\end{array}
\right.
\]
Hence, the Dirac-delta distribution of free-space $S$ {\em maximizes} $n_{\infty}(p)$ over the entire range $0 \leq p \leq 1$.
}
\end{thm}

These findings suggest that under the equal flow-redistribution model considered here, power systems with homogeneous distribution of redundant space are  
more robust against random attacks and failures, as compared to systems with heterogeneous distribution of redundancy. Interestingly, this suggests that the optimal robustness is achieved when the tolerance factor $\alpha_i = S_i/L_i$ decreases with increasing load, leading to the counter-intuitive conclusion that the lines carrying the highest load should have the smallest tolerance factor to achieve maximum robustness; see \cite{yingrui_yagan_optimal} for a more detailed discussion on this matter.

\begin{figure}[!t]
  \vspace{-2mm}
  \centering
  \includegraphics[width=.4\textwidth]{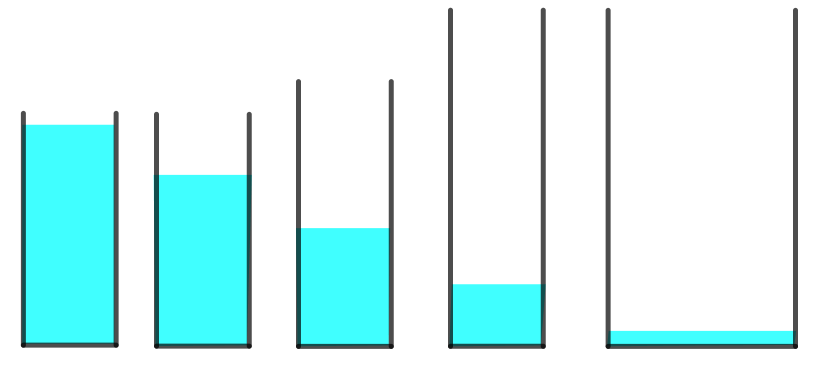}
  \vspace{-1mm}\caption{ \sl In this example  we have (load, capacity) values given by $(8,8+\epsilon), (6,8+\epsilon), (4,4+{14}/{3}+\epsilon), (2,11+\epsilon), (1,21+\epsilon)$ where $\epsilon>0$ is arbitrarily small. The  greedy maximum-load attack will need to attack $k=5$ containers to fail all. It will start attacking the leftmost container with load $L_1=8$ which will not lead to any further failures. Then, it will continue with the second one from the left, again unable to trigger a cascade, and continue until attacking all containers directly. The optimal solution can be seen to be $k=1$ by attacking the last container, which will trigger a cascading failure  destroying the whole system. We can generalize this counterexample to the case with $N$ containers with the greedy algorithm's output being $k=N$ while the optimal solution being $k=1$. \vspace{-4mm}}
  \label{fig:counter_example_max_load}
\end{figure}

\section{The optimal attack problem and insights}
\label{sec:attack_perspective}
While existing results in \cite{yingrui_yagan_optimal,yagan_fiber_bundle} shed some light on the robustness under the equal flow redistribution model, many real-world threats to power systems would be expected to come from powerful adversaries targeting specific parts of the grid to inflict maximum damage on the overall system. In order to understand the system's robustness against such {\em targeted attacks} and to reveal its most vulnerable lines, we now consider the problem from an adversary's perspective and seek effective strategies for attacking the system under given constraints. In particular, we consider a scenario where the adversary has full information about the system and aims to find the best set of $k$ of lines to attack so that it fails maximum number of lines as a result cascading failures. This optimization problem is formally introduced next.

\subsection{The Main Optimization Problem: ER-$k$}
The {Equal Redistribution} (ER) problem with $k$ attacks is the {\em optimization problem}, denoted ER-$k$, that aims to find the set $A$ of $k$ lines such that attacking $A$ leads to the maximum number of total line failures (as a result of load redistribution and cascading failures), among all possible attack sets with size $k$. Put differently, we seek to find $A$ with $|A|=k$ that minimizes $n_{\infty}(A)$. Throughout, we find it useful to consider  the {\em decision} version of this optimization problem (referred to as the ER-$k$-$k'$ problem) formally defined as follows.

\vspace{1mm}
\textbf{INPUT:} $N$ pairs of non-negative numbers in the form $(L_i,C_i)$ indicating the load and the capacity of each line, and integers $k$ and $k'$ such that $0<k<k'\le N$. We assume $C_i > L_i$ so that no line fails initially at its own load. 

\vspace{1mm}
\textbf{OUTPUT:} The answer to whether or not there is an attack set $A$ with size $k$, such that at the end of the cascading failures the number of failed nodes is at least $k'$; i.e., whether there exists $A$ with $|A|=k$ and $n_{\infty}(A) \leq N-k'$.

\subsection{Heuristic Algorithms that Fail}
Here we will present three intuitive greedy algorithms and give concrete examples demonstrating their poor performance for the optimization problem described above. In doing so, we will focus on the special case where $k'=N$ meaning that the goal of the attack is to destroy the whole system, by attacking a minimum number $k$ of lines. 

In what follows, we find it useful to describe the problem in a simpler way, where we have $N$ water containers with capacities $C_1, \ldots, C_N$, and initial water levels 
 $L_1, \ldots, L_N$. As in the equal flow-redistribution model, when a container is "attacked"  its content is redistributed equally to the remaining containers. Also, if the water level in a container exceeds its capacity, we assume that it has failed and redistribute its content, again equally, to the remaining containers. With this formulation, the goal of the attackers is to find the smallest number $k$ of containers to target so that all containers get overloaded and fail eventually.

An important observation is that the following intuitive algorithms can deviate significantly from the optimal solution.
\paragraph{Greedy max-load attack} This greedy strategy aims to maximize the load that will be redistributed in each attack round. Namely, it starts by attacking the container with the highest load,  and proceeds similarly, waiting after every attack for a steady-state to be reached (meaning that all load redistribution and potential further failures end). The algorithm stops when all containers fail. This strategy is not optimal in general because it fails to recognize the opportunity to eliminate containers with very large capacities that will otherwise be difficult to fail by redistributing the load.
The worst-case deviation from the optimal (in terms of the number of lines needed to be attacked for complete system failure) is $\Theta(n)$; e.g., see Figure \ref{fig:counter_example_max_load}.

\begin{figure}[!t]
 \vspace{-2mm}
  \centering
  \includegraphics[width=.35\textwidth]{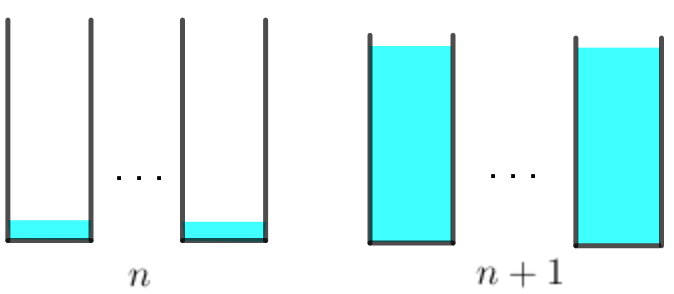}
 \vspace{-2mm} \caption{\sl Consider $2n+1$ containers where (load, capacity) 
  values are given by $(\epsilon, M)$ for the first $n$ containers and $(M-2\epsilon,M-\epsilon)$ for the last $n+1$ containers; here $\epsilon>0$ is arbitrarily small
and $M>2(n+1)\epsilon$. 
The greedy max-capacity attack will need to attack $k=n+1$ containers to fail the all containers; it will start attacking the first $n$ containers but cascading failures will not take place. On the other hand, the optimal solution is $k=1$ as it takes to attack only one of the containers with $(M-2\epsilon,M-\epsilon)$ to trigger a cascading failure that will fail all.} \vspace{-4mm}
  \label{fig:counter_example_max_capacity}
\end{figure}

\paragraph{Greedy max-capacity attack} This strategy is similar in spirit with the greedy max-load attack except that this time the container with the maximum capacity is attacked in each round. The idea here is that by taking out large containers, the remaining, supposedly small, containers will be destroyed due to load redistribution. This strategy is not in general optimal either, because there may be containers with {\em large} capacities but small (or, even almost zero) loads, rendering an attack to such containers very ineffective in terms of triggering failures by means of load redistribution.
The worst-case deviation from the optimal is  again $\Theta(N)$ as demonstrated in Figure \ref{fig:counter_example_max_capacity}.

\begin{figure}[!t]
\vspace{-4mm}
  \centering
  \includegraphics[width=.34\textwidth]{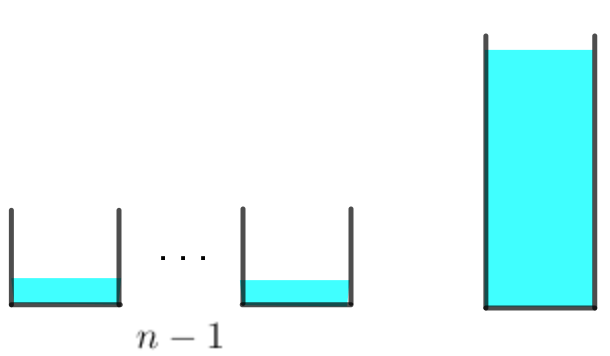}
  \vspace{-2mm}\caption{ \sl In this example we have $n$ containers with
  (load, capacity) values 
   $(\epsilon, (n+1)\epsilon)$ for the first $n-1$ containers and $(M,M+(n-1)\epsilon)$ for the last container, where $\epsilon>0$ is arbitrarily small and $M$ satisfies $M>(n^2-n)\epsilon$. The greedy max free-space ($C-L$) attack will output $k=n$ since it will start attacking the leftmost containers and no cascading failures will take place. The optimal solution is obviously $k=1$ by attacking the last container.}\vspace{-4mm}
  \label{fig:counter_example_max_space}
\end{figure}

\paragraph{Greedy max-free-space attack} 
It is clear from the previous two cases that the optimal attack strategy will be one that 
considers both the loads and capacities of the containers involved. 
The greedy approach that targets containers with largest free space (i.e., $(capacity - load)$ difference) 
falls into this category, and is based on the fact that containers with largest free space will fail the {\em latest}
in the course of a cascading failure; e.g., see Section \ref{subsec:observations} for a discussion of this fact. 
Therefore, it is sensible to eliminate those containers with a direct attack.
On the other hand, containers with small free space are already on the verge of failing and therefore can be taken down by means of
redistribution of loads. Although this greedy strategy is intuitive (and in fact optimal in some special cases), it fails to be the optimal solution in general. The main reason is that this approach does not take into account the loads of the containers directly. For example, a container may have a large free space but its load may be negligible compared to other containers, rendering a direct attack on this container ineffective. The worst-case deviation from the optimal is again $\Theta(N)$  as demonstrated in Figure \ref{fig:counter_example_max_space}.

\subsection{Observations towards Designing a Smart Algorithm} 
\label{subsec:observations}

We now present some observations that will be useful in designing a {\em smart} attack algorithm.

\paragraph{The order of attack does not matter} In the equal redistribution model, the order with which we launch an attack does not affect the final set of failed containers. This is because the load of the attacked nodes will be distributed to all of the remaining nodes so at the end an amount of $\sum_{i \in A}L_i$ will end up in the remaining containers (leading to new failures or not) irrespective of the order we chose to  attack the containers in $A$. We remark that an attack strategy can still be designed in a greedy fashion, where the set $A$ is determined one member at a time, waiting for cascades to stop after each attack. 

\paragraph{Order of failures during the cascading process} Assume that containers are sorted by increasing free space, $S_i = (C_i - L_i)$.
Given that any failed load is redistributed {\em equally} among the remaining containers, it is clear that this ordering will remain the same throughout the course of cascading failures; the containers that are attacked directly at the beginning are excluded from this argument.
Therefore, in the process of recursive load redistribution, containers will fail (due to their free space diminishing to (below) zero) in this exact same order: the one with smallest free space will fail first, and so on and so forth.

\section{Optimal Attack Strategies under Special Cases} 
\label{sec:greedy_success}
We now present three special cases of the ER-$k$ problem and provide {\em optimal} attack strategies for each of them.


\paragraph{Same Loads}
An interesting situation arises when initial loads are the same for all containers while capacities differ. This reflects situations in which all lines in the power system are given the same initial load, but have different capacities owing to the physical constraints or material used. We show  that a greedy algorithm finds the optimal solution in this special case.
The ER-$k$-Same Loads Problem is defined formally as follows.
\vspace{1mm}

\noindent \textbf{INPUT}: A non-negative rational number $L$ for the common load and a list of $N$ non-negative numbers $C_i> L, \forall i$ indicating the capacity of each line. The integer $k$ represents the number of attacks we can launch.
\vspace{1mm}

\noindent \textbf{OUTPUT}:  The set $A$ of lines to be attacked   that minimizes $n_{\infty}(A)$ under the constraint $|A|=k$.
\vspace{1mm}

The next result, proved in  Supplementary Material, shows that the max-$C$-greedy algorithm finds the optimal solution.

\begin{thm} The $max$-$C$-Greedy Algorithm is optimal for the ER-$k$-Same Loads Problem.
\label{thm:same_load_optimal}
\end{thm}

\paragraph{Same Free Spaces}
\label{subsec:same_space}
Sometimes it might be the case that the containers have arbitrary load and capacity but they have a fixed free space. 
In \cite{yingrui_yagan_optimal}, this was in fact shown to be the optimal design that gives maximum robustness against {\em random} attacks.
We refer to the corresponding problem as the ER-$k$-Same Free Spaces, formally defined as follows. 


\vspace{1mm}
\noindent
\textbf{INPUT}: A list of $N$ non-negative rational numbers $L_i$ indicating the load of each container and a positive rational number $S$ indicating the common free space.

\vspace{1mm}
\noindent\textbf{OUTPUT}: Find the minimum number $k$ of containers needed to be attacked in order to destroy the whole system.
\vspace{1mm}

We changed the output  from having a fixed number of lines to be attacked to inflict the maximum damage, to the case where we aim to  destroy the whole system with the minimum number of attacks. This is because in the case where every container has the same free space, 
there are no intermediate cascading failures. After an attack, the system will either fail completely, or no single line will fail other than those attacked directly. 
We show in Supplementary Material that the $max$-$L$-Greedy algorithm that targets lines with the largest loads leads to the optimal solution for this problem.
\begin{thm} 
{\sl The $max$-$L$-Greedy Algorithm is optimal for the ER-$k$-Same Free Spaces Problem.}
\label{thm:max_L_opt}
\end{thm}

\paragraph{Capacities Proportional to Loads}

In many cases, the capacities and the loads of power lines  are related in a particular way. Namely, the capacity of a line is often set to be proportional to its load. For example with $\alpha>0$ denoting the {\em tolerance factor}, we have
$C_i = (1+\alpha) L_i$ for each line $i=1,\ldots, N$.
In this variation, we will also show that there is a greedy algorithm achieving the optimal solution.
The ER-$k$-$(C \propto L)$ Problem is defined formally as follows.

\vspace{1mm}
\noindent\textbf{INPUT}: A list of $N$ non-negative  numbers $L_i$ indicating the load of each container and a positive number $\alpha$ such that container capacities are set to $C_i = (1+\alpha)L_i$ for each $i$. 

\vspace{1mm}
\noindent\textbf{OUTPUT}:  The set $A$ of lines to be attacked   that minimizes $n_{\infty}(A)$ under the constraint $|A|=k$.

In this setting, load, free-space, and capacity of the lines will be ordered in the same way, and as we show in the Supplementary Material, $max$-$L,C,S$-Greedy algorithms that target lines with the largest load {\em and} free-space {\em and} capacity give the optimal solution to this problem. 
\begin{thm}
{\sl The $max$-$L,C,S$-Greedy Algorithms are optimal for the ER-$k$-$(C \propto L)$ Problem.
}
\label{thm:3_osy}
\end{thm}


\section{A modified optimal attack problem with total load constraints}
\label{sec:hardness}

In this Section, we will prove that a variation of the decision problem ER-$k$-$k'$ is NP-Complete. In particular, we consider the ER-$k$-$k'$-$ Q$ problem, defined formally as follows.

\textbf{INPUT:} $N$ pairs of non-negative numbers in the form $(L_i,C_i)$ indicating the load and the capacity of each line, integers $k$ and $k'$ such that $0<k<k'\le N$,
and a positive number $Q$.  We also assume $C_i > L_i$ for each $i=1,\ldots, N$.

\textbf{OUTPUT:} The answer to whether or not there is an attack set $A$ with size  $k$, and total sum of loads $\sum_{i \in A} L_i\le Q$, such that at the end of the cascading failures the number of failed nodes is at least $k'$; i.e., whether there exists $A \subset \{1,\ldots, N\}$ with $|A|=k$,  $\sum_{i \in A} L_i\le Q$, and $n_{\infty}(A) \leq N-k'$.

It is clear that the objective is two-fold here and that there is an inherent trade-off: by attacking lines with larger initial loads we can shed more load on other lines and have a better chance to trigger a cascade of failures that would destroy the whole system. However, the problem enforces a constraint on the total load of the attacked containers as well. This {\em knapsack-like} trade-off is what makes the problem NP-complete as we now show.
Our proof is based on the reduction of the ER-$k$-$k'$-$Q$ problem from the {\em $k$-Subset Sum} variant defined as follows:  {\em Given a set of integers and a target sum $Q$, is there any subset of size ${k}$ whose sum is $Q$?}

\begin{thm} 
{\sl The ER-$k$-$k'$-$Q$ Problem is NP-Complete.}
\label{thm:NP_Hard_result}
\end{thm}
\myproof
First, we show that  ER-$k$-$k'$-$Q$ Problem is in NP: The certificate is a list of the $k$ containers we choose to attack. We can check in polynomial time (e.g., see the ER-Attack Projection algorithm in \cite{vaggos_ITA}) whether at least $k'$ lines in the system fail or not. Since we have a certificate that can be checked in polynomial time, ER-$k$-$k'$-$Q$ is in NP!

Given an instance of the $k$-Subset Sum problem we will create an instance of the ER-$k$-$k'$-$Q$ problem: Given a set of $N$ integers $a_1,a_2,...,a_N$, the $k$-Subset Sum problem asks whether there exists $k$ members of the set whose some equals $Q$. If $k=N$, we can check if $\sum_{i=1}^N a_i = Q$ and respond accordingly. From now on, we suppose $k<N$ and create an equivalent version of the ER-$k$-$k'$-$Q$ problem in the following manner. Let lines $\mathcal{L}_1, \ldots, \mathcal{L}_N$ have loads $L_1=a_1,L_2=a_2,...,L_N=a_N$ and consider the 
ER-$k$-$k'$-$Q$ problem; i.e., we seek to find a set $A$ of $k$ lines such that $\sum_{i \in A} L_i \leq Q$ and that attacking $A$ leads to failure of at least $k'>k$ lines in the system. We also set $C_i=L_i+S_i$ where the free space is $S_i=\frac{Q}{N-k}$ for each $i=1,\ldots, N$. 
This last constraint ensures two things. First, as discussed in Section \ref{sec:greedy_success}, when all lines have the same free space then attacking $k$ lines can only have two consequences: either only those $k$ lines that are attacked fail, or all $N$ lines fail. In either case, there is no {\em cascade} of failures 
and the system reaches a steady-state immediately.
Thus, with equal free space among all lines, the ER-$k$-$k'$-$Q$ problem with $k' > k$ is equivalent to
ER-$k$-$N$-$Q$ problem.  Secondly, under the enforced assumptions it is clear that a complete system failure will take place if and only if the total load failed by the initial attack $A$ is larger than the sum of the free spaces of those that are not in the attack set $A$; i.e., if and only if
\vspace{-2mm}
\[
\sum_{i \in A} L_i \geq \sum_{j \in \{1,\ldots,N\}/A} S_j = (N-k) \frac{Q}{N-k} = Q.
\]
Here, the first equality follows from the facts that $|A|=k$ and $S_i=\frac{Q}{N-k}$ for each $i=1,\ldots, N$. 
Recalling further the constraint that $\sum_{i \in A} L_i \leq Q$, this leads to $\sum_{i \in A} L_i = Q$. Therefore, the created instance of the ER-$k$-$k'$-$Q$ problem indeed seeks to find a subset $A$ of $\{a_1, \ldots, a_N\}$ such that $|A|=k$ and $\sum_{i \in A} L_i = Q$, rendering it equivalent to the $k$-Subset Sum instance that we have started with. 
For the reverse direction, assume that the ER-$k$-$k'$-$Q$ problem has a solution with $k$ lines $\mathcal{L}^{(1)}, \ldots, \mathcal{L}^{(k)}$. Then the loads of these lines constitute a solution to the $k$-Subset-Sum problem.

The above reduction can be constructed in polynomial time (more precisely, in linear time), so if there was a polynomial algorithm that could solve the  ER-$k$-$k'$-$Q$, then the $k$-Subset Sum would be in P, which is wrong unless P=NP. Thus, we conclude that the  ER-$k$-$k'$-$Q$ Problem is NP-complete.
\myendpf
\vspace{-2mm}

An important implication of the above result is that the {\em optimization} version of the  ER-$k$-$k'$-$Q$ problem, which seeks to find the set $A$ of lines  that minimizes $n_{\infty}(A)$
under the constraints $|A|=k$ and $\sum_{i \in A} L_i \leq Q$, is NP-Hard. This means that under these constraints, the adversary can not launch an optimal attack in polynomial time unless P=NP. 


\section{Heuristic algorithms and their performance}
\label{heuristic}
Although, it is not known whether the original optimization problem of finding the best $k$ lines to attack to minimize final system size is NP-Hard or not, the discussion in the preceding section indicates that the optimal attack problem is likely to be computationally challenging; in particular, we know that the problem is NP-Hard if we are further constrained by the total load of those we can attack. This prompts us to develop heuristic algorithms, for both the original and the modified optimization problems, that work in polynomial time and have competitive performance under arbitrary load-capacity distributions. The performance of these heuristics will then be compared with some benchmark heuristics such as max-$L$, max-$C$, max-$S$, and random attacks. 

In the interest of brevity and concreteness, the discussion is restricted here to {\em non-greedy} algorithms that choose the attack set $A$ {\em without} ever running the attack projection algorithm. In other words, the algorithms are not allowed to run the cascading failures initiated by a subset of $A$, and then continue making the remaining selections from the lines that survived the cascade. Of course, all heuristics considered here including the benchmarks can be modified to operate in a greedy fashion. 
One might also be tempted to use a greedy algorithm where, at each round, the line to be attacked is chosen in an optimal way; i.e., the line whose failure leads to smallest number of surviving lines is chosen from among all lines that are still alive. However, for the problem at hand, one can realize that unless $|A|$ is relatively large, the final system size equals $n_{\infty}(A) = N-|A|$ regardless of the set $A$ of attacked lines. For example,  with $|A|=k$, this will be the case whenever
\[
k \leq \frac{S_{min}}{L_{max}} (N-k),
\]
meaning that if  ${S_{min}}/{L_{max}}>0$,  the greedy heuristic will have to deal with {\em ties} when making its choices for the next line to be attacked until it makes
$\Omega(n)$ choices. Since resolving the ties by randomization for such a large number of selections is likely to lead to poor performance, one needs heuristic rules to resolve the ties.
Even then, our preliminary simulation study indicated that the greedy versions of the heuristics considered here perform only slightly better than their non-greedy counterparts, and the comparison among the greedy heuristics provided no additional insight to what was already observed from non-greedy algorithms; hence the decision to consider only non-greedy attack strategies here.

\subsection{Heuristics for the original optimization problem}
\label{heuristic_no_constraint}

We first consider the original case where there is no constraint on the total load of the lines that can be attacked; i.e., we consider the ER-$k$ problem.  
Let $A$ be the set of lines to be attacked such that $|A|=k$. It is clear from the previous discussions that a good attack should aim to 
\begin{align}
&\textrm{maximize} ~\sum_{i \in A} L_i, ~~ \textrm{and}
\label{eq:ER_k_obj1}\\
&\textrm{minimize} ~ S_j, ~ j \in \{1,\ldots, N\}-A
\label{eq:ER_k_obj2}
\end{align}
In words, the attack should aim to find the lines with the largest free space while making the total load of the failed lines as large as possible. Thus, the attack should intuitively look for lines with {\em large} initial load {\em and} {\em large} free-space. Of course, most difficult situations arise when the load and free-space values of the lines are in reverse order; e.g., the highest load carrying line has the smallest free-space, etc. as in Figure \ref{fig:counter_example_max_load}.

Our main idea towards handling the trade-off described above is based on its similarities with the well-studied 0-1 Knapsack problem. In the 0-1 Knapsack problem, we are given a set of $N$ items, $\{1,\ldots, N\}$, each with a weight $w_i$ and a value $v_i$, and the goal is to choose items 
such that their total value is maximized while the total weight is bounded by $W$; i.e.,
\begin{align}
&\textrm{maximize} \sum_{i} v_i x_i
\label{eq:Knapsack_obj}\\
&\textrm{subject to} \sum_{i} w_i x_i \leq W \quad \textrm{and} \quad x_i \in \{0,1\} 
\label{eq:Knapsack_contraints}
\end{align}
The 0-1 Knapsack problem is known to be NP-Hard, but polynomial-time heuristics can still give close-to-optimal solutions. For example, a competitive heuristic is to order the items based on their  "value per weight", i.e., $v_i/w_i$, and choose items according to this order, starting with the one with the highest $v_i/w_i$,  until the total weight capacity $W$ is reached. In fact, with a small modification to handle corner cases, this heuristic is known to yield at least $50 \%$ of the optimum value. 

The optimal attack problem we consider, i.e., the ER-$k$ problem, has some similarities with but is not equivalent to the 0-1 Knapsack problem. In particular, one can construct an analogy between the constraints of the 0-1 Knapsack problem and the ER-$k$ problem by assigning all item weights as $w_i=1$ and the total weight limit to be $W=k$. However, the objectives (\ref{eq:ER_k_obj1})-(\ref{eq:ER_k_obj2}) of the ER-$k$ problem are much more complex than the objective (\ref{eq:Knapsack_obj}) of the Knapsack problem. Nevertheless, the two problems have some similarity in that their main difficulties lie in the trade-offs involved. In the Knapsack problem the trade-off is between the value and the weight of the item and it is desirable to pick items with high value and low weight, while in the ER-$k$, the trade-off lies between the possibly conflicting objectives of choosing lines with high load and high free-space. Inspired with the efficient heuristic  for the Knapsack problem that is based on selecting items with the largest $v_i/w_i$ ratio (i.e., items with the biggest {\em bang for the buck}), our first heuristic for the ER-$k$ problem is based choosing lines with the highest load times free-space, i.e., with the highest $L_i*S_i$.

\paragraph{Maximum Load$\times$Free Space Attack} In this
algorithm, the load free space product, $L_i * S_i$ is computed for each
line $i=1,\ldots, N$. After sorting the lines based on this product, the
$k$ lines to be attacked is chosen as the ones having the
highest $k$ load-free space product. As mentioned above, this is inspired by the 2-approximation heuristic for the Knapsack problem that orders items according to $v_i * \frac{1}{w_i}$ when the goal is to choose items with high $v$ and low $w$. In the ER-$k$ problem, we wish to choose lines with high $L$ and high $S$, or equivalently, with high $L$ and low $\frac{1}{S}$. Thus, constructing an analogy between value $v_i$ and load $L_i$, and weight (or, cost) $w_i$ and $1/S_i$, our heuristic chooses lines with the maximum
\[
L_i * \frac{1}{1/S_i} = L_i * S_i.
\]
The performance of this heuristic is demonstrated via several numerical examples in the next subsection along with a comparison with some benchmark heuristics. 

Aside from its connection to a powerful heuristic in a relevant problem, the maximum  $L*S$ heuristic has several advantages. First of all, this heuristic becomes equivalent to the {\em optimal} attack strategy in the three special cases considered in Section \ref{sec:greedy_success}; e.g., when all lines have the same load, it chooses ones with highest free-space (and hence capacity), or when all lines have the same free-space, it chooses lines with maximum $L$, etc. Secondly, considering the product $L*S$ is an effective way to favor lines with high load {\em and} free-space, while heavily penalizing load or free-space values close to zero;
note that benchmark heuristics including the highest-capacity attack ($C=L+S$) fail to penalize 
small $L$ or $C$ values.
In the optimal attack problem, this makes perfect sense given that a line with almost no load should never be attacked even if it has very high free-space since its failure will likely not affect any other line. Similarly, it may not be a good idea to directly attack a line with almost no free-space even if it has very high load, since the line will likely fail due to load redistribution {\em regardless} of which other lines are attacked.

\paragraph{Maximum $L * S^{\beta}$ attack} While maximum $L*S$ attack is intuitive and will be seen to be powerful in many cases, we observe that its performance can be further improved by a small modification. To this end, we propose a second heuristic as a modified version of the max-$L*S$ attack that allows adjusting the relative importance of load and free-space values of the lines. In particular, with $\beta$ in $[0,\infty]$, we consider a heuristic that chooses $k$ lines with the maximum 
$
L_i * S_i^{\beta}$. 
An added benefit of this heuristic is that it contains several heuristics as special cases. In particular, the maximum $L*S$
algorithm described above is obviously a special case of this algorithm, corresponding to the case $\beta=1$. Also, by setting $\beta=0$, this heuristic reduces to the max-$L$ attack, while setting $\beta=\infty$ (or, large enough) makes it equal to the max-$S$ attack.

\subsection{Numerical Comparison with Benchmark Heuristics}
We  now compare the heuristics we developed against some benchmark heuristic algorithms via numerical experiments. The benchmark heuristics we will consider are given below:

\paragraph{Random attack} This is the most primitive attack strategy and considered here only for comparison purposes. The attack picks $k$ lines to be attacked uniformly at random from amongst all $N$ lines.

\paragraph{Highest-$L$, highest-$C$, highest-$S$ attacks} These three attacks are based on sorting lines with
respect to their initial load $L_i$, free-space $S_i$, or capacity $C_i=L_i+S_i$, respectively and attacking the top $k$ lines with the highest value of the corresponding metric. 





We fix the number of lines at $N=5000$ for all experiments. 
First, we consider the case where each line is independently given an initial load from a uniform distribution, $U(L_{min},L_{max})$, where we set $L_{min}=10$ and $L_{max}=30$. The free-space allocated to each line is generated independently from its load, again from  a uniform distribution, $U(S_{min},S_{max})$ with $S_{min}=10$, $S_{max}=60$. The capacity of a line $\mathcal{L}_i$ is given by the sum $C_i=L_i+S_i$. The independence of $L$ and $S$ leads to some lines having high load  but small free-space, or vice versa, rendering the optimal attack problem {\em non-trivial}; e.g., with these choices, the realized load-capacity values will almost surely {\em not} fall into one of the special cases presented in Section \ref{sec:greedy_success} where an optimal solution is available.

\begin{figure}[!t]
\vspace{-3mm}
  \centering
  \includegraphics[width=.49\textwidth]{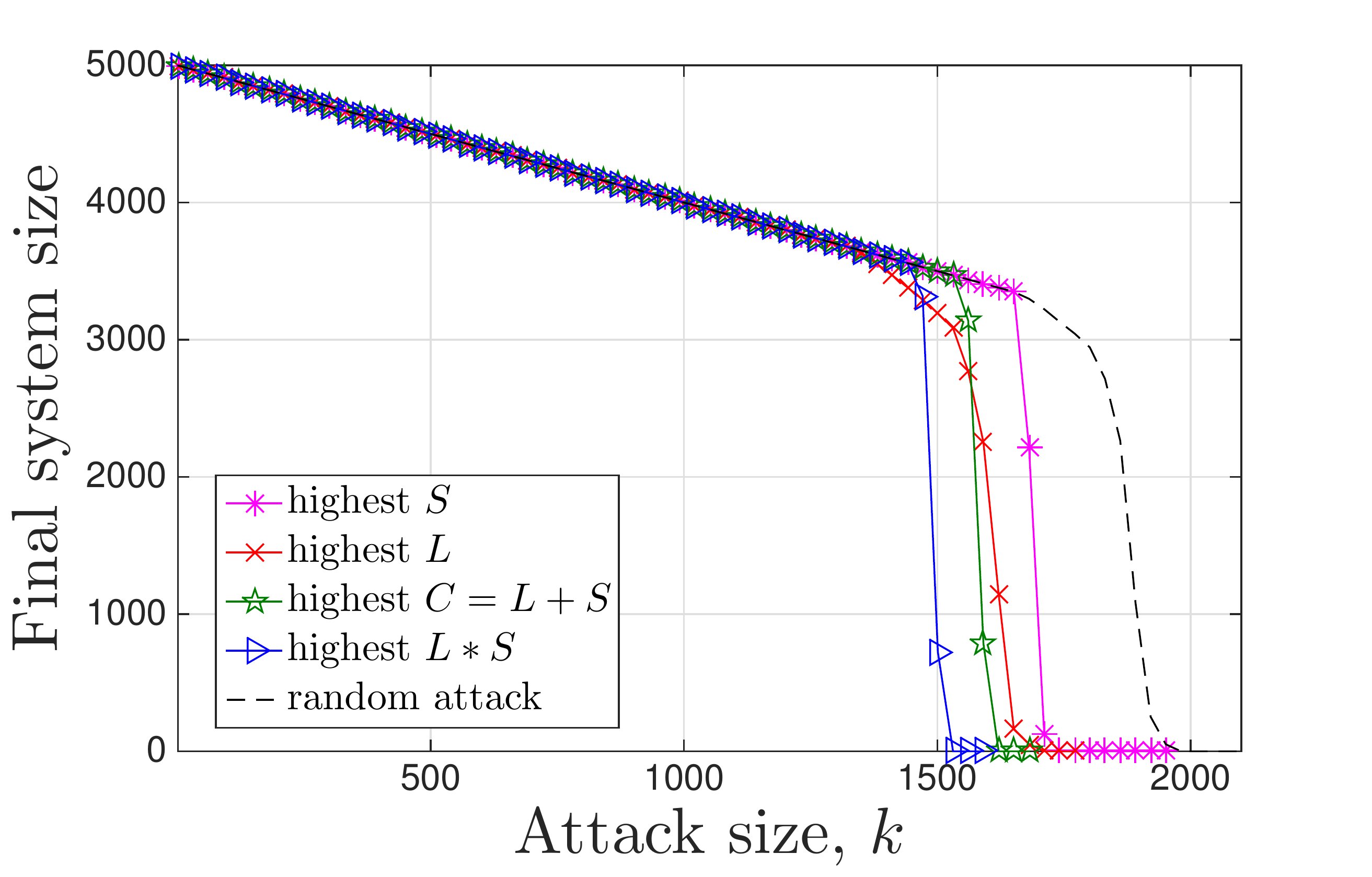}
 \vspace{-5mm} \caption{ \sl The performance comparison of different heuristic algorithms for $L\sim U[10,30]$,
  $S\sim U[10,60]$, $N=5000.$}
  \label{fig:L_10_30_S_10_60}\vspace{-4mm}
\end{figure}

Under this setup, we compute the final system
size as a function of the number $k$ of lines attacked, where the set of attacked lines are selected according to the heuristics considered. The results are given in  Figure~\ref{fig:L_10_30_S_10_60} where each data point is obtained by averaging over 100 independent runs. 
We already see in this simple setting that our attack strategy of targeting lines with the highest $L*S$ outperforms all other benchmarks (except a small interval of attack size where max-$L$ attack seems to give the highest damage). In particular, we see that the highest $L*S$ attack is able to fail the whole system by targeting $90$, $180$, $210$, and $450$ fewer lines as compared to max-$C$, max-$L$, max-$S$, and random attacks, respectively.

\begin{figure}[!t]
\vspace{-1mm}
  \centering
  \includegraphics[width=.52\textwidth]{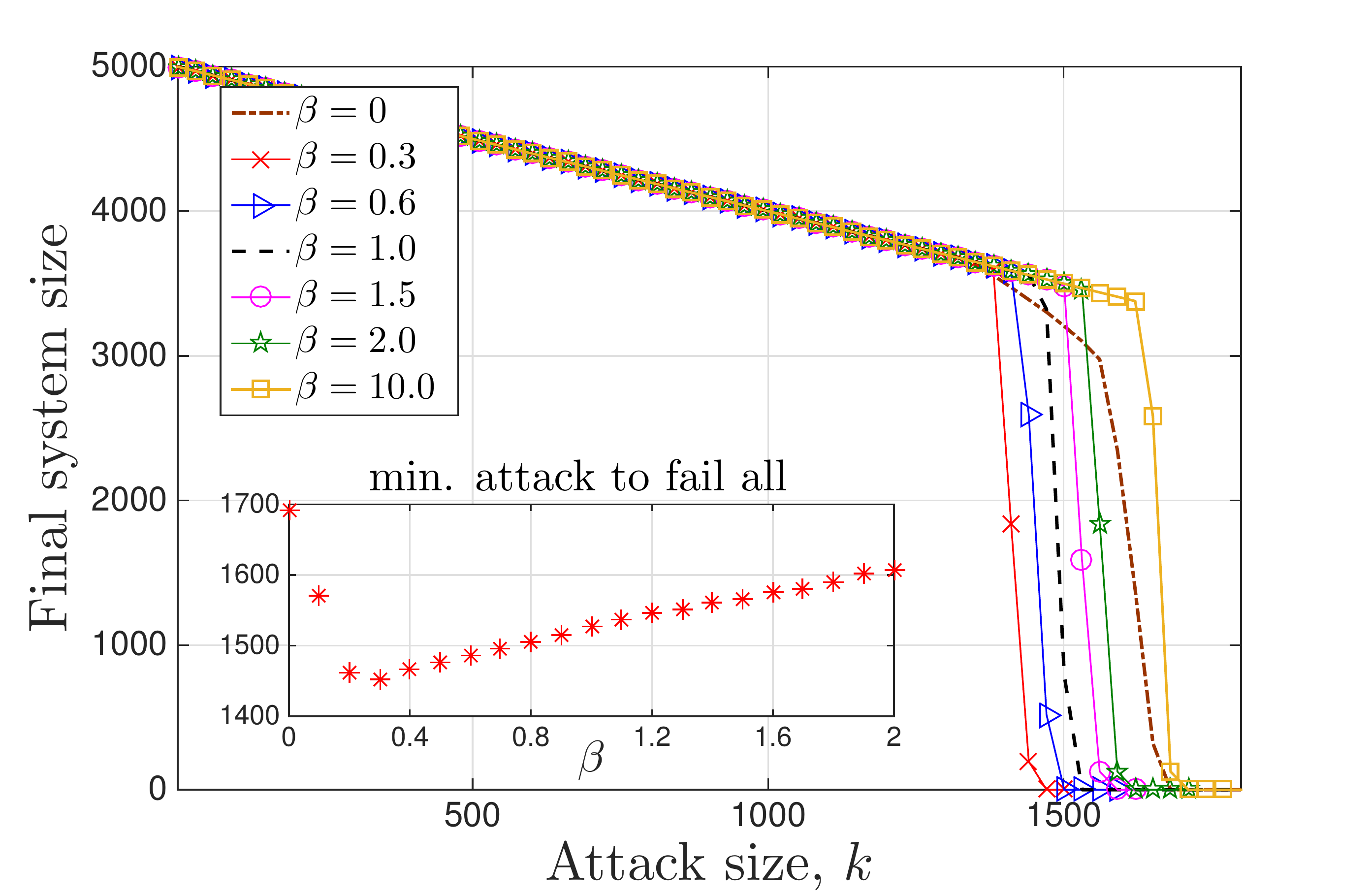}
 \vspace{-6mm}  \caption{{\em The performance comparison of maximum $L * S^\beta$ algorithms for various $\beta$ values for $L\sim U[10,30]$,
  $S\sim U[10,60]$, $N=5000.$ Inset: The minimum number of lines needed to be attacked to fail all lines for the maximum $L* S^{\beta}$ attack.}}\vspace{-3mm}
  \label{fig:beta_0_one_half}
\end{figure}

Next, we check if this performance can be further improved by attacking lines with highest $L*S^{\beta}$ for some $\beta \geq 0$. To this end, we repeat the previous experiment as $\beta$ varies from zero to ten. The results are demonstrated in Figure \ref{fig:beta_0_one_half} and as expected show that with $\beta=0$ or $\beta \gg 1$, we obtain the same performance with max-$L$ and max-$S$ attack, respectively. More interestingly, we see that the case $\beta=1$ is indeed {\em not} the best one can do. For example, we see that when $\beta=0.3$, the max-$L*S^{\beta}$ attack can fail the whole system by attacking 75 fewer lines the case for $\beta=1$. To demonstrate this better, we plot in the inset of Figure \ref{fig:beta_0_one_half} the minimum number lines needed to be attacked to fail all $N$ lines. To compute this, we again run 100 independent experiments and pick the minimum attack size for which {\em all} 100 experiments led to entire system failure. 

The performance of heuristic algorithms are known to vary significantly under different parameter settings, and our case is not expected to constitute an exception. To that end, we have tested the performance of the max-$L*S^{\beta}$ attack with $\beta \in [0,4]$, for a wide range of possibilities for the distribution of $L$ and $S$. In all cases we considered, we were able to identify a $\beta$ value for which the max-$L*S^{\beta}$ attack was at least as good as all benchmark attacks (random, highest-$C,L,S$-attacks) showing its versatile performance. 

As already mentioned, the most challenging cases arise when the load and free-space values are in reverse order. To that end, we close this section by demonstrating the performance of the max-$L*S^{\beta}$ attack in such cases. In particular, we start by generating $L_1,\ldots, L_N$ and $S_1,\ldots, S_N$ independently according to some distribution. Then, the load values (resp.~free-space values) are sorted and re-arranged in increasing (resp.~decreasing) order, leading to highest-load carrying line having the smallest free-space, and so on. To make the problem more challenging and interesting, we also consider {\em Pareto} distribution. Namely, a random variable $X$ is said to follow Pareto distribution, written $X \sim \textrm{Pareto}(X_{\textrm{min}}, b)$ with
$X_{\textrm{min}}>0$ and $b>0$, if its probability density is given by
\[
p_X(x) =  X_{\textrm{min}}^{b} b x^{-b-1} \1{x \geq X_{\textrm{min}}}.
\]
To ensure that $\bE{X}=b X_{\textrm{min}}/(b-1)$ is finite, one must set $b>1$, while the variance of $X$ is finite only if $b > 2$.

\begin{figure}[!t]
\vspace{-4mm}
  \hspace{-2mm}
  \includegraphics[width=.5\textwidth]{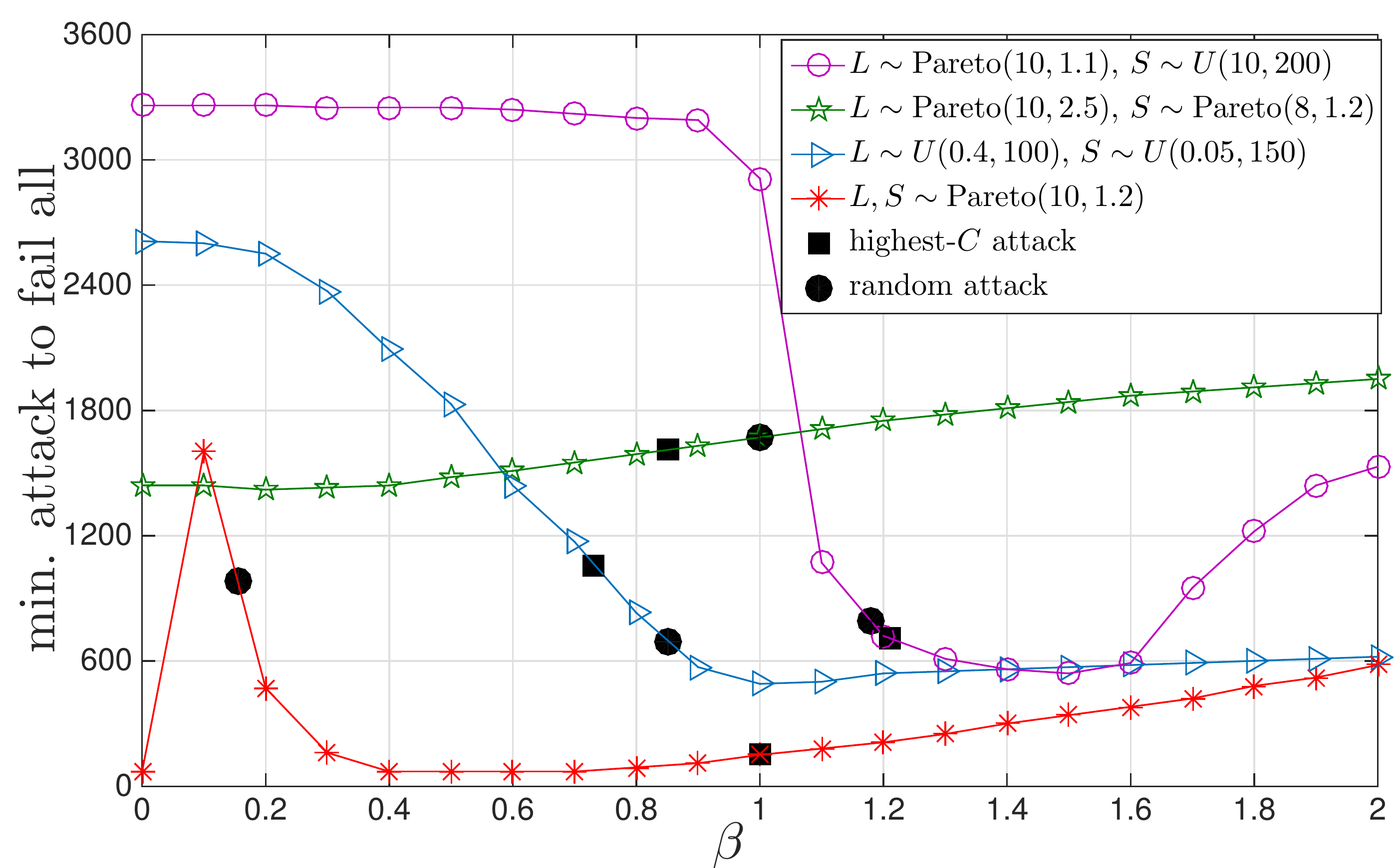}
  \vspace{-5mm} \caption{ {\sl  Minimum number of lines needed to be attacked to fail all lines in the system is shown when load and free-space values are generated independently (from the distributions given at the figure legend) and then sorted in reverse orders; e.g., to ensure $L_1 \leq L_2 \leq \cdots \leq L_N$, while $S_1 \geq S_2 \geq \cdots \geq S_N$. Curves stand for the results obtained under the max-$L*S^{\beta}$ attack as a function of $\beta$. Corresponding results for the max-$C$ attack are shown by  filled square symbols, and those for the random attack by filled circles.}}
  \label{fig:reverse_order}\vspace{-2mm}
\end{figure}

The results for the case where $L$ and $S$ values are reverse ordered are depicted in Figure \ref{fig:reverse_order}. Here, we show a small number of representative results that correspond to different behaviors for brevity. As before, all results correspond to the minimum attack size that led to an entire system collapse in all 100 experiments. The curves represent the results for the max-$L*S^{\beta}$ attack as $\beta$ varies from zero to two. In each plot, we add the corresponding results for the max-$C$ attack (shown by a filled square) and random attack (shown by a filled circle) as well; for convenience, the $x$-axis values for these symbols are chosen such that they stay on the corresponding curve showing the results for max-$L*S^{\beta}$ attack.
We note that the max-$L$ attack is already demonstrated by the case $\beta=0$ while $\beta=2$ gives a good indication of the performance of the max-$S$ attack, so these plots provide a comparison of the max-$L*S^{\beta}$ attack with all the benchmarks considered here.

The main observations from Figure \ref{fig:reverse_order} are as follows. We see that in all cases there is a particular $\beta$ value for which the  max-$L*S^{\beta}$ attack performs the best among all benchmarks; it is only the case of $L,S\sim \textrm{Pareto}(10,1.2)$ where we see that the best
performance of max-$L*S^{\beta}$ attack is attained when $\beta=0$, or $\beta \in (0.4, 0.7)$ meaning that max-$L$ attack matches the performance of the max-$L*S^{\beta}$ attack. Also, we see that the $\beta$ value that leads to the best performance can be equal to, smaller than, or larger than one in different scenarios showing the importance of trying different values of $\beta$ to get the best performance. Finally,  while benchmark attacks (including the random attack) occasionally give results close to the best max-$L*S^{\beta}$ attack, we see examples for each benchmark where its performance is significantly worse than the best max-$L*S^{\beta}$ attack; these cases are summarized in Table
I.

\begin{table}[!b]
 \label{tab:performance}
\begin{tabular}{|c||c|c|c|c|c|}
  \hline Distribution of &  \multicolumn{5}{c|}{Minimum \# of lines to attack to fail all} \\ \cline{2-6}
$L$ and $S$ &   random & max-$C$  & max-$L$ & max-$S$
 & best $\beta$
\\
\hline \hline
 $L \sim \textrm{Pareto}(10,1.2)$ 
 & & & & & \\ 
 $S \sim \textrm{Pareto}(10,1.2)$ 
 & \boldsymbol{$981$}   & \boldsymbol{$151$}  & 71  &  \boldsymbol{$2241$} & 71 \\ \hline 
 $L \sim U(0.4, 100)$  & & & & & \\ 
 $S \sim U(0.05, 150)$ & $691$    & \boldsymbol{$1061$} & \boldsymbol{$2611$}   & \boldsymbol{$1021$}  & 491 \\\hline 
 $L\sim \textrm{Pareto}(10,2.5)$ & & & & & \\  $S\sim \textrm{Pareto}(8,1.2)$ & $1671$    & $1611$ & $1421$   & $2111$  & 1411 \\ \hline 
 $L \sim \textrm{Pareto}(10,1.1)$
  & & & & & \\ 
   $S \sim U(10,200)$
 & \boldsymbol{$791$}    & $711$ & \boldsymbol{$3261$}   & \boldsymbol{$2221$}  & 541 \\
 \hline 
  $L, S$ from the 
  & & & & & \\ 
    UK National Grid$^{\ast}$ 
 & \boldsymbol{$2371$}    & $1491$ & $1611$   & \boldsymbol{$2111$}  & 1441 \\
 \hline 
 \end{tabular}\vspace{2mm}
\caption{\sl Performance comparison of benchmark attacks with the best result of the max-$L*S^{\beta}$ attack. The first four rows are obtained from Figure \ref{fig:reverse_order}, while the last row is obtained from  simulations with UK National Grid data (see Section \ref{sec:UKdata} for details).
Values  significantly worse (in the sense of needing to attack many more lines to fail all) than the best-$L*S^{\beta}$ attack are made bold.}
\vspace{-.3cm}
\end{table}

\subsection{Heuristic attacks for the modified optimization problem}

We now consider the modified optimization problem ER-$k$-$k'$-$Q$ where the attack set $A$ is further constrained by $\sum_{i \in A} L_i \leq Q$, in addition to $|A| \leq k$. As shown in Theorem \ref{thm:NP_Hard_result}, in this case finding the attack set $A$ that minimizes the final system size is NP-Hard, prompting us to develop heuristic strategies.
With the additional constraint on the total load of the lines that we can attack, the trade-offs involved become more complicated and heuristics developed in the previous section may not be well-suited for the ER-$k$-$k'$-$Q$ problem. Ultimately, our strategy should be to choose an attack set that has $k$ (or, very close to $k$) lines with total load equal (or, very close) to $Q$,
and that have the highest free space among all lines. This is because at the first stage of the cascades, any line $\mathcal{L}_i$ that was not directly attacked will fail only if 
\[
S_i \leq \frac{\sum_{i \in A} L_i}{N-|A|}.
\]
Thus, to facilitate failures it is desirable to make $\sum_{i \in A} L_i$ and $|A|$ as large as possible, while $S_i$ as small as possible. 

Given the multiple constraints involved, this problems shows similarity with the {\em 2-dimensional 0-1 Knapsack problem} \cite{loulou1979new,senju1968approach,toyoda1975simplified}:
Consider a collection of items, where each item $i$ is given a value $v_i$, has weight $w_i$, and volume $q_i$. The objective is then to maximize $\sum_{i}v_i x_i$ subject to $\sum_{i} w_i x_i \leq W$ {\em and}
$\sum_{i} q_i x_i \leq Q$ where  $x_i \in \{0,1\}$; i.e., we want to choose items with the maximum total value while the total weight is limited by $W$ and total volume is limited by $Q$. As can be inferred from the discussion above, an important difference of the ER-$k$-$k'$-$Q$ problem is that while it is desirable to choose lines with high $S_i$ (could be thought to be analogous to the {\lq\lq value\rq\rq} of the item) under given constraints, it is perhaps equally important to attain or be very close to the limits on both total load and total number of lines attacked. 

\begin{figure*}[t] 
    \centering
\subfigure[]
    {
    \includegraphics[width=0.46\linewidth]{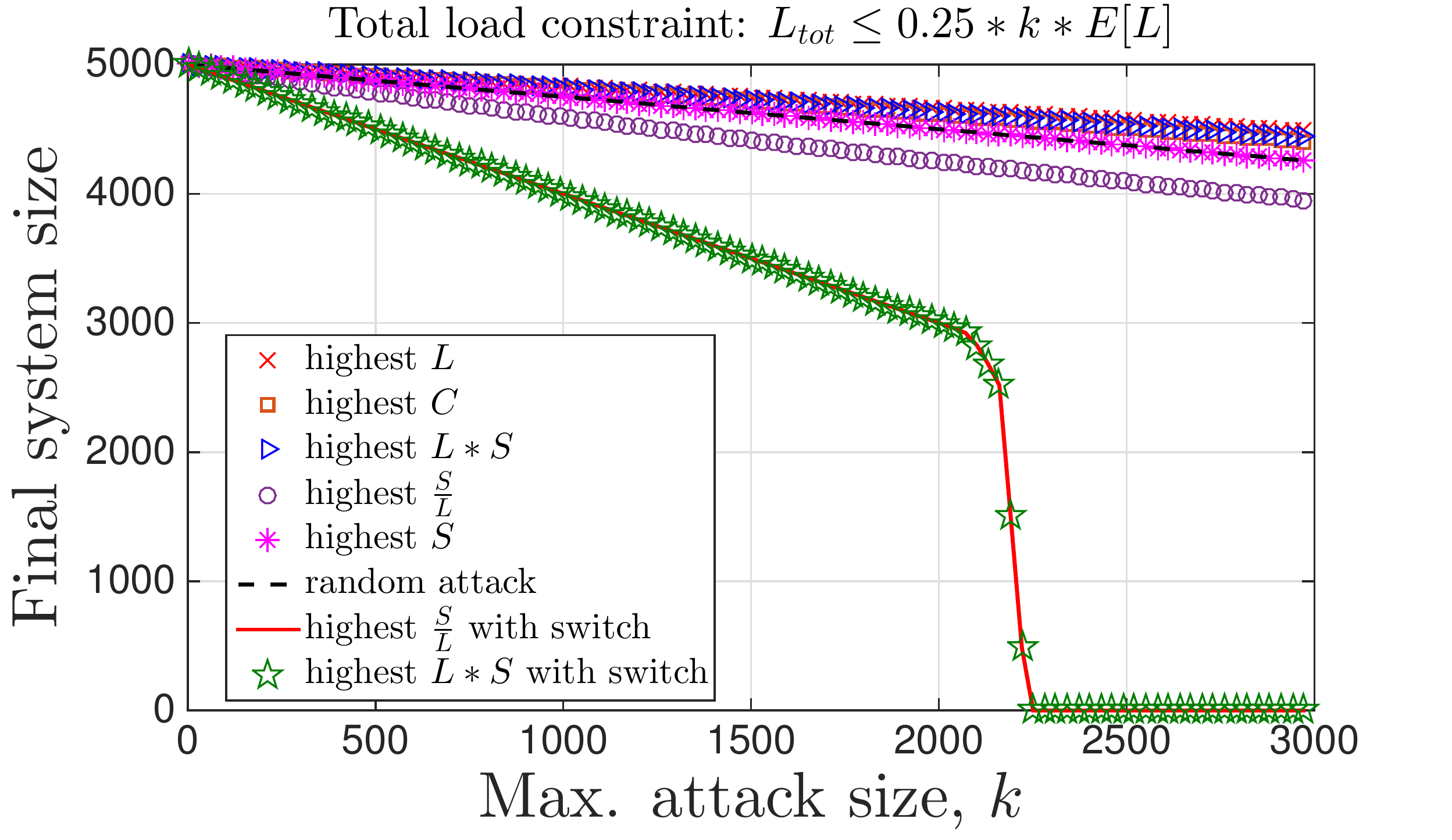} 
    \label{fig:last_indep_1}} 
\subfigure[]
    {
    \includegraphics[width=0.46\linewidth]{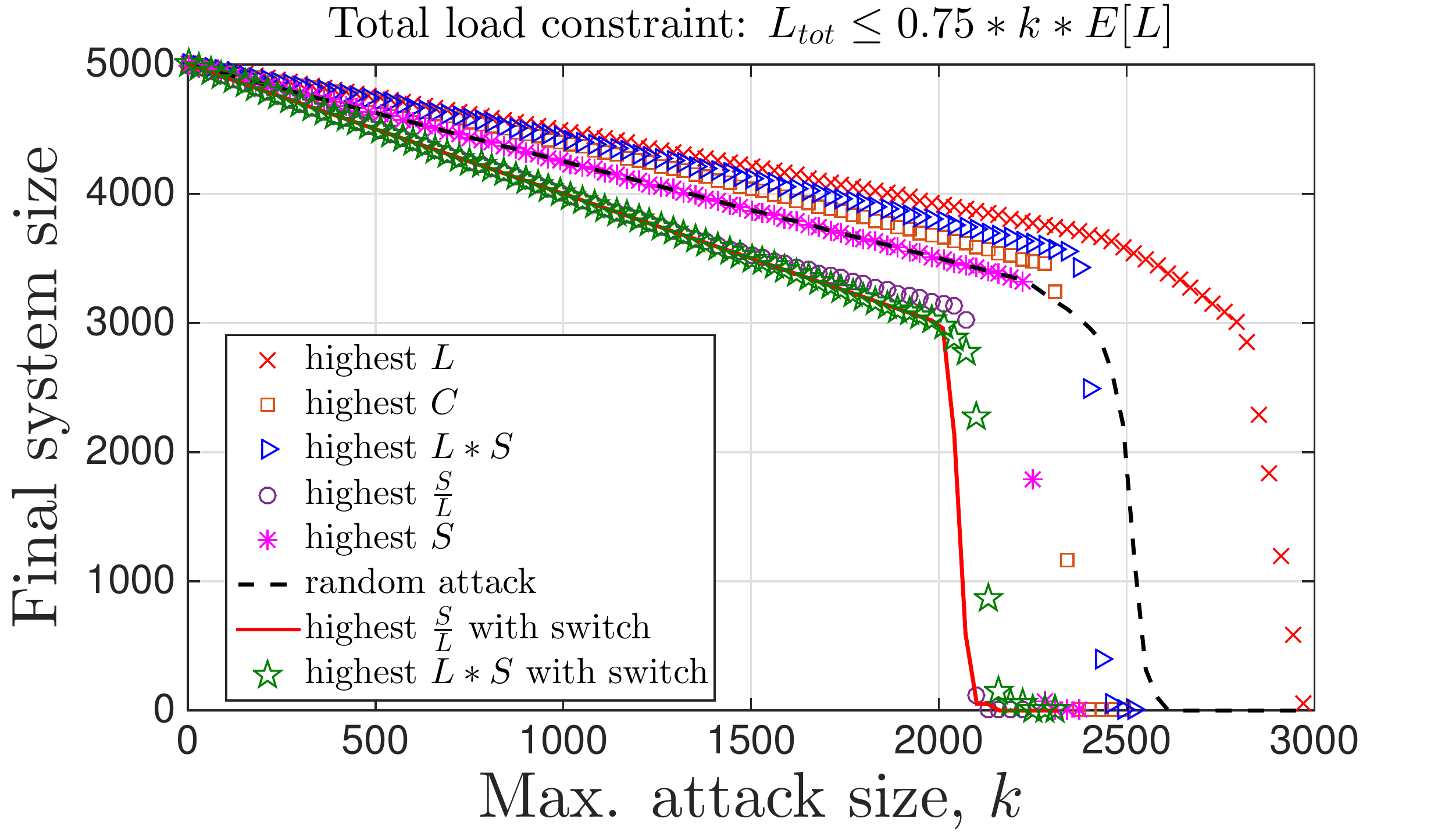} 
    \label{fig:last_indep_2}} 
    \subfigure[]
    {
    \includegraphics[width=0.46\linewidth]{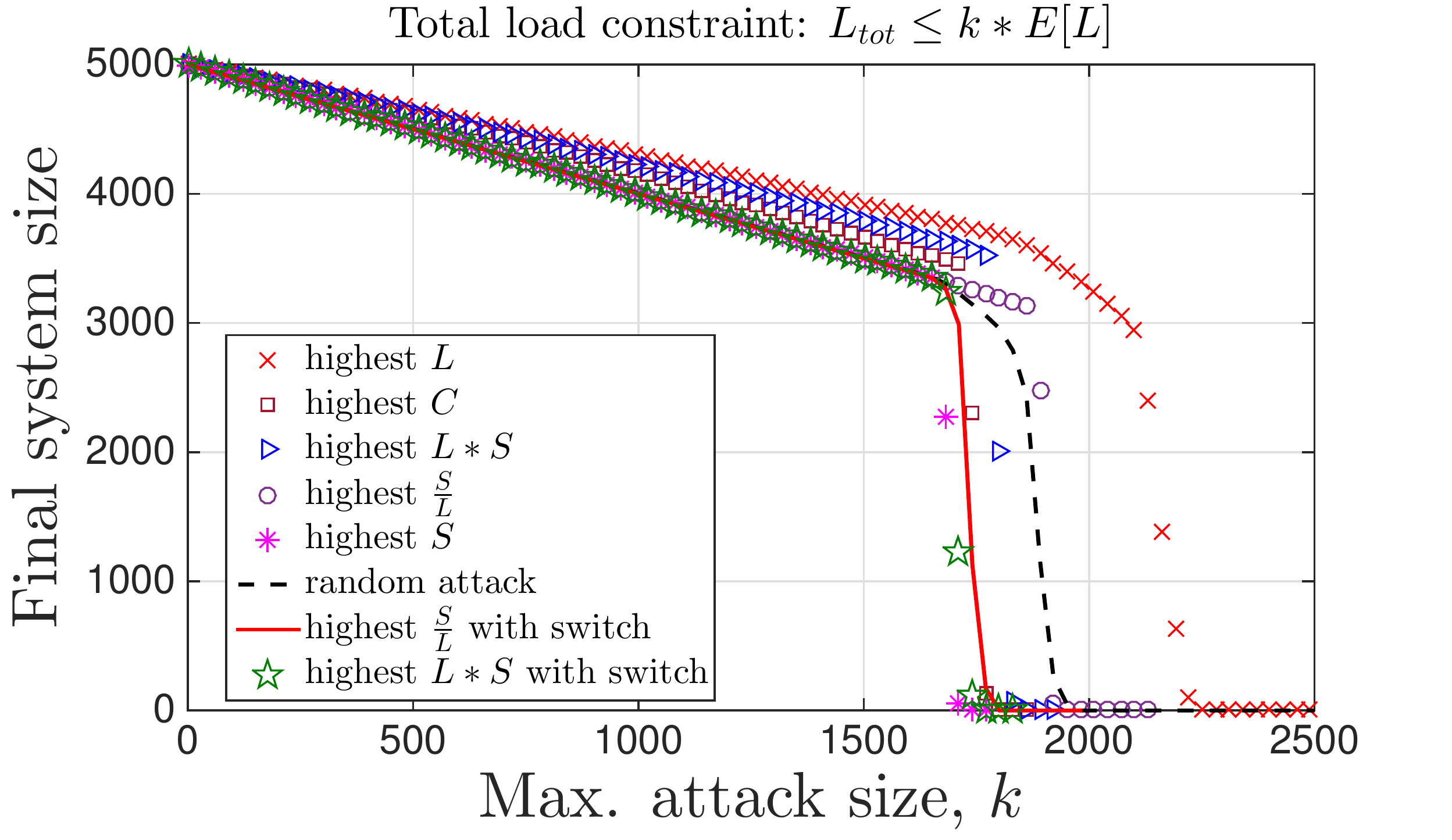} 
    \label{fig:last_indep_3}} 
    \subfigure[]
    {
    \includegraphics[width=0.46\linewidth]{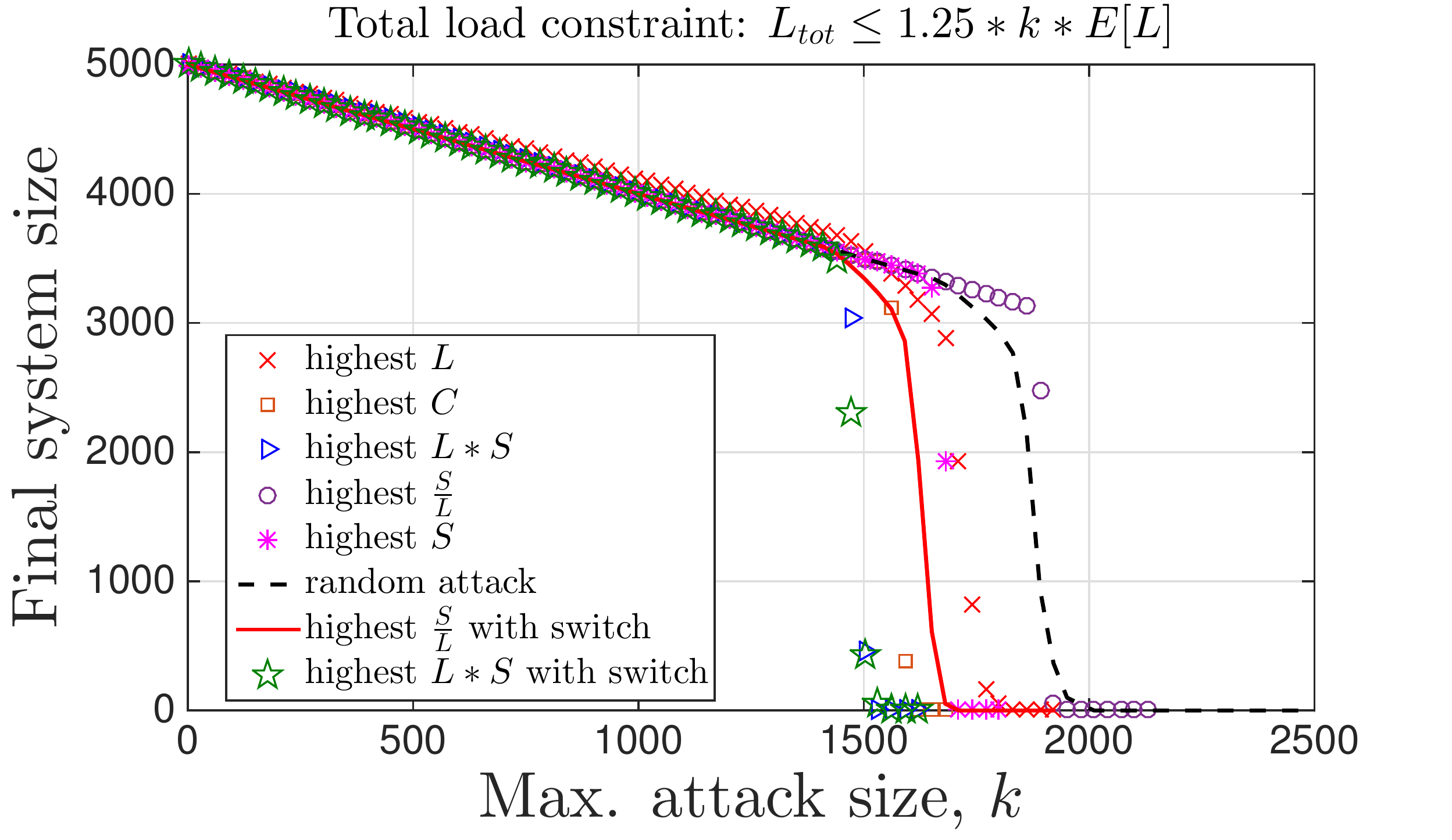} 
    \label{fig:last_indep_4}} 
    \hspace{8mm}
 \vspace{-4mm} \caption{ \sl The performance comparison of different heuristic algorithms for $L\sim U[10,30]$,
  $S\sim U[10,60]$, $N=5000$, when the attack is constrained to $k$ lines such that their total load satisfies a) $L_{\textrm{tot}} \leq 0.25*k*\mathbb{E}[L]$; b) $L_{\textrm{tot}} \leq 0.75*k*\mathbb{E}[L]$; c) $L_{\textrm{tot}} \leq 1.0*k*\mathbb{E}[L]$; d) $L_{\textrm{tot}} \leq 1.25*k*\mathbb{E}[L]$. \vspace{-3mm}}
  \label{fig:last_indep} 
\end{figure*}

With these in mind, our heuristics for the ER-$k$-$k'$-$Q$ problem are based on incorporating a {\em switch} to the previously developed heuristics
that is actuated to ensure that the attack set $A$
attains or gets close to both constraints on its cardinality and the total load.
This idea is inspired from the greedy-like heuristic  developed for the multi-dimensional Knapsack problem in \cite{loulou1979new}. This algorithm initially starts choosing items   based on a given set of rules until one or more of the constrained resources become scarce, and then switches to a different set of rules that favor items that use very little of the scarce resource. Here, we propose to use a heuristic that chooses the lines to be attacked one at a time according to the previously developed max-$L*S$ strategy. After each selection, we check whether the switch needs to be activated. Namely, with $A'$ denoting the set of $k'$ lines selected so far, we check
\begin{itemize}
\item[i)] Is it still feasible to select all remaining $k-k'$ lines from the smallest load carrying lines available? Namely, with the remaining lines' loads sorted in ascending order $L^{(1)} \leq L^{(2)} \leq \cdots  \leq L^{(N-k')}$, we check if
\vspace{-1mm}
\[
\sum_{j=1}^{k-k'}L^{(j)} + \sum_{i \in A'} L_i {\leq}^{?} Q
\]
If the answer is YES, we continue with the second condition for the switch, while if the answer is NO, the switch is activated and algorithm is finished by appending $A'$ with $k-k'-1$ lines with the smallest load-carrying lines available; i.e., with $\mathcal{L}^{(1)},\ldots \mathcal{L}^{(k-k'-1)}$. Alternatively, one can release the latest added member of $A'$ and append it with $k-k'$ lines $\mathcal{L}^{(1)},\ldots \mathcal{L}^{(k-k')}$; we found no major performance difference between these two approaches.
\item[ii)] Next,  check whether it is feasible to select all remaining $k-k'$ lines from the largest load carrying lines available. Namely, with the remaining lines' loads sorted in ascending order $L^{(1)} \leq L^{(2)} \leq \cdots  \leq L^{(N-k')}$, we check if
\[
\sum_{N-k+1}^{N-k'}L^{(j)} + \sum_{i \in A'} L_i {\leq}^{?} Q
\]
If the answer is NO, we continue the algorithm with the next selection, while if the answer is YES, the switch is activated and algorithm is finished by appending $A'$ with $k-k'$ lines with the largest load-carrying lines available; i.e., with $\mathcal{L}^{(1)},\ldots \mathcal{L}^{(k-k'-1)}$. 
\end{itemize}

The two conditions of the switch described above ensure that while the initial selections are made in line with the original objectives of picking lines with high load and high free-space, care is also given so as to be able to pick $k$ (or, $k-1$) lines whose total load is close to the limit $Q$. Of course, any algorithm including the benchmarks can be modified using the switch idea to better accommodate total load constraints. In particular, when the total load limit $Q$ is extremely stringent, it would be tempting to pick lines with small load so as to not exhaust the total load limit quickly, while aiming to choose lines with high free-space. This prompts us to consider the max-$S/L$ heuristic as well, including its modification with the {\em switch} idea described above. To keep the discussion brief we do not present results for the max-$L*S^{\beta}$ attack (with or without switch) and consider only the case where $\beta=1$; this is in part due to the fact that when the switch is added, the performance of the max-$L*S^{\beta}$ attack becomes much less sensitive to variations in $\beta$ over small ranges.

\begin{figure*}[t] 
\vspace{-2mm}
    \centering
\subfigure[]
    {
    \includegraphics[width=0.46\linewidth]{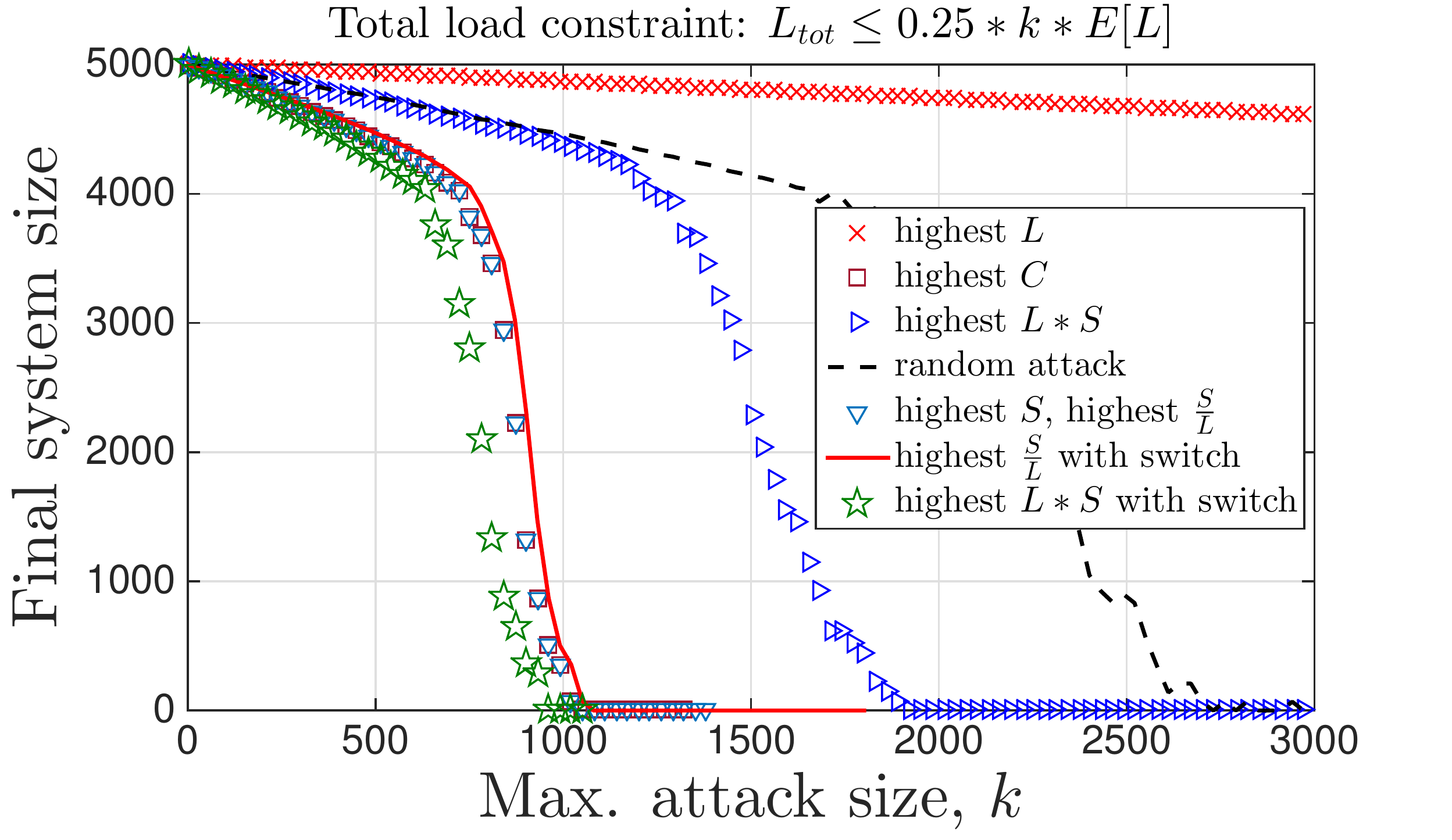} 
    \label{fig:last_inverse_1}} 
\subfigure[]
    {
    \includegraphics[width=0.46\linewidth]{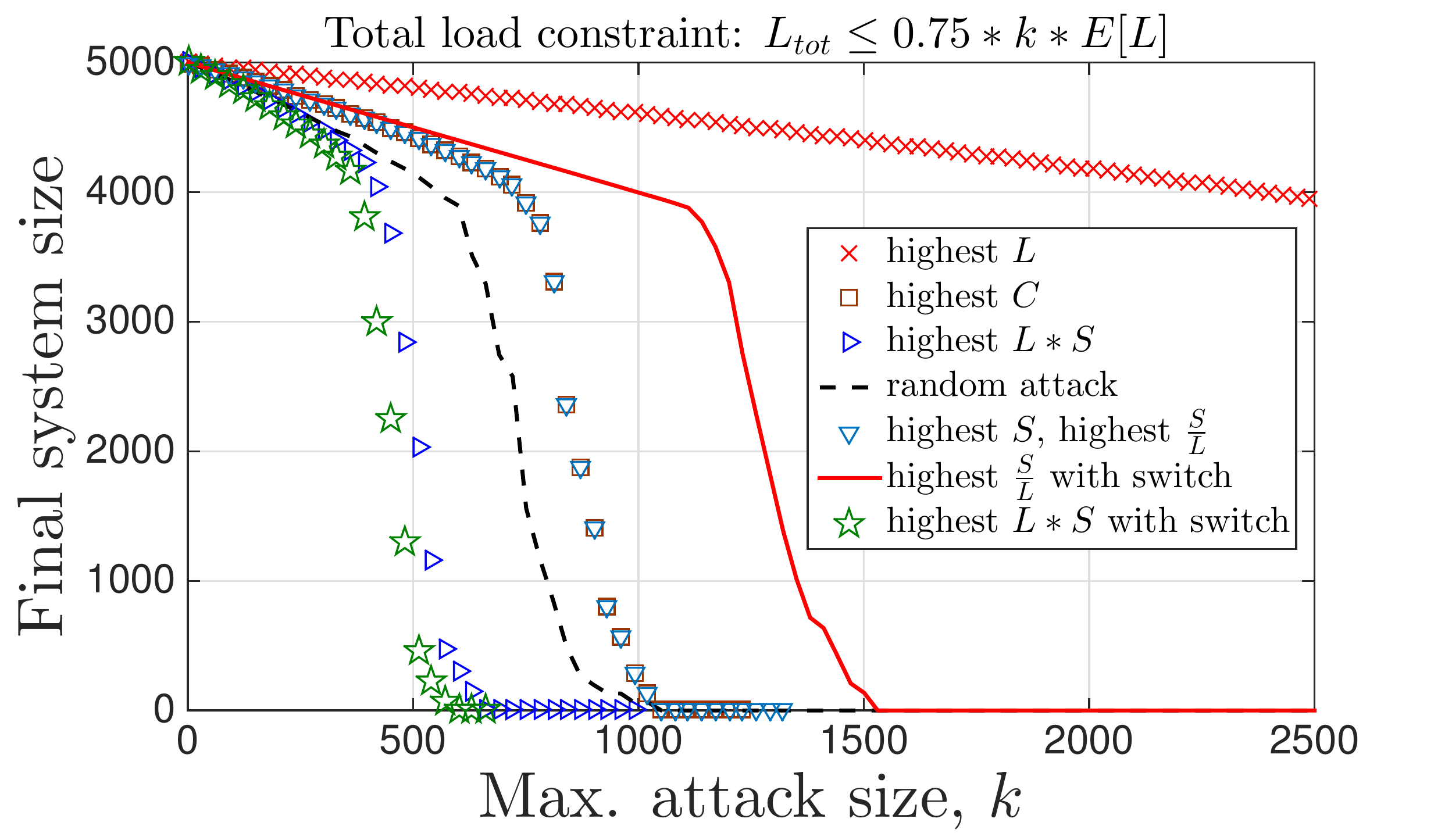} 
    \label{fig:last_inverse_2}} 
    \subfigure[]
    {
    \includegraphics[width=0.46\linewidth]{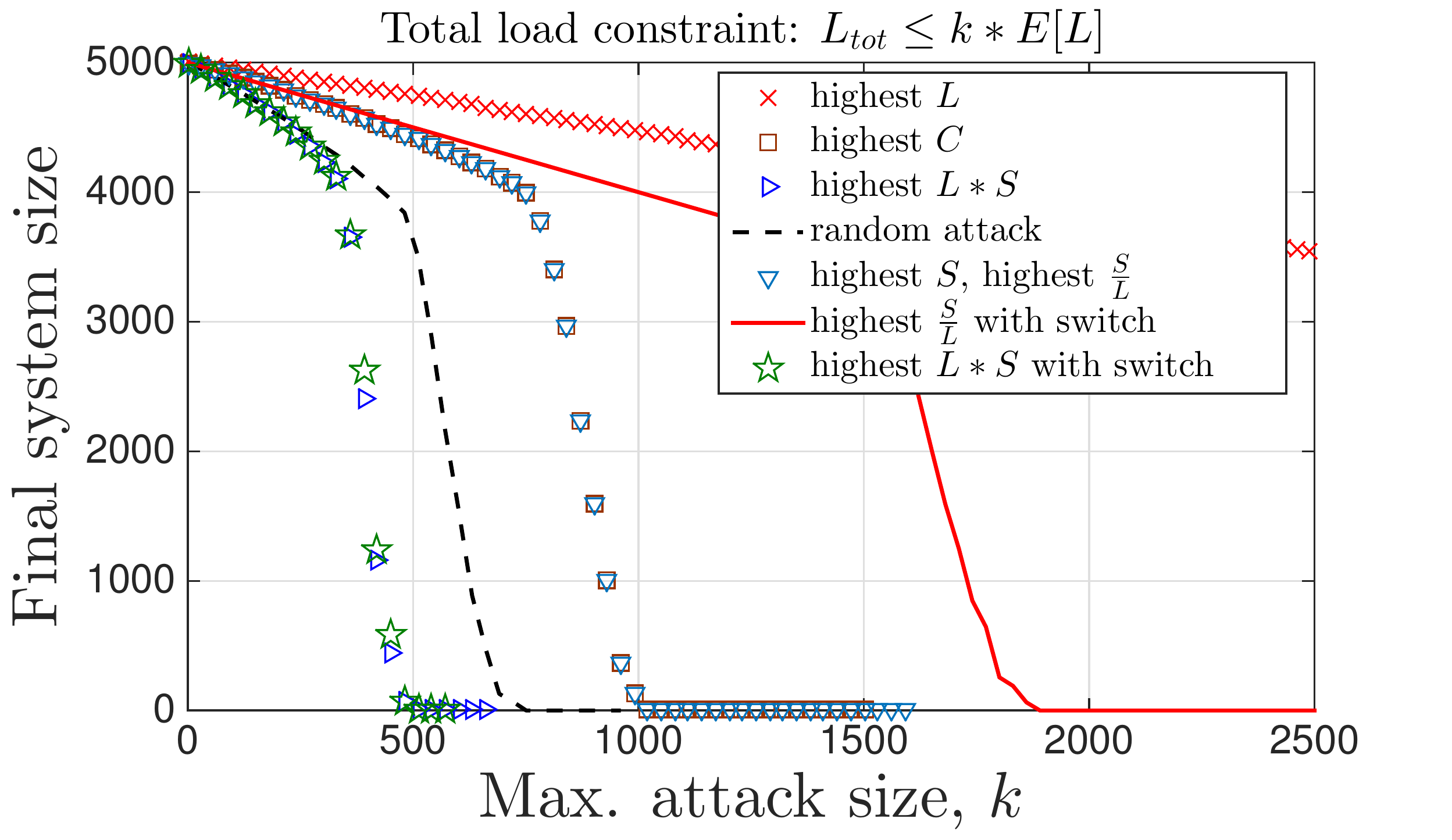} 
    \label{fig:last_inverse_3}} 
    \subfigure[]
    {
    \includegraphics[width=0.46\linewidth]{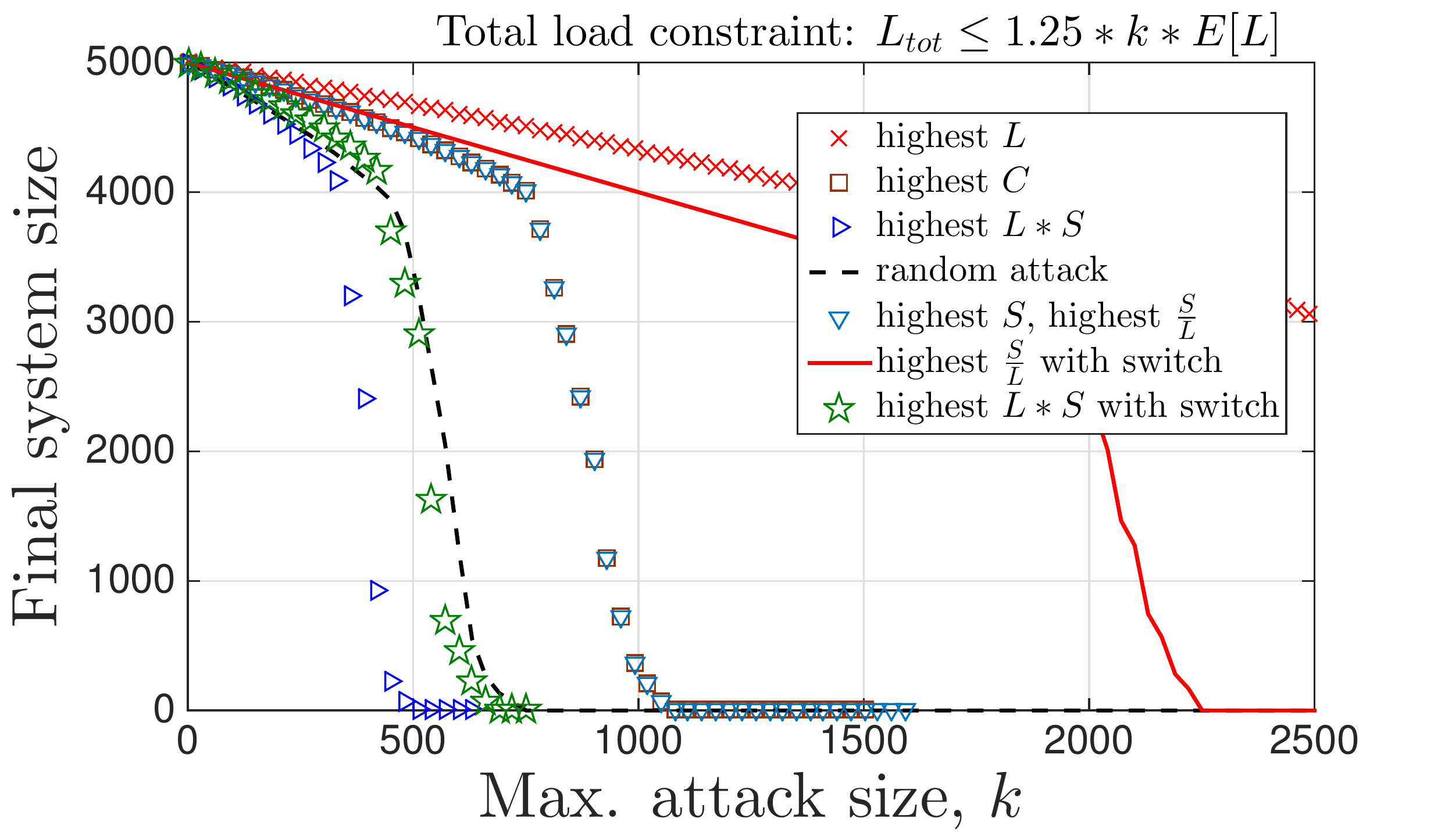} 
    \label{fig:last_inverse_4}} 
    \hspace{8mm}
\vspace{-4mm} \caption{ \sl The performance comparison of different heuristic algorithms for $L\sim U[0.4,100]$,
  $S\sim U[0.05,150]$, with $L$ and $S$ sorted in reverse order, $N=5000$, when the attack is constrained to $k$ lines such that their total load satisfies a) $L_{\textrm{tot}} \leq 0.25*k*\mathbb{E}[L]$; b) $L_{\textrm{tot}} \leq 0.75*k*\mathbb{E}[L]$; c) $L_{\textrm{tot}} \leq 1.0*k*\mathbb{E}[L]$; d) $L_{\textrm{tot}} \leq 1.25*k*\mathbb{E}[L]$. Each data point is obtained by averaging over 100 independent runs.}
  \label{fig:last_inverse} \vspace{-4mm}
\end{figure*}

We now present numerical results to evaluate the performance of the heuristic attacks developed here and compare them against benchmark algorithms; as before, we will use max-$C,S,L$ and random attacks as benchmarks. Different from the experiments conducted in 
Section~\ref{heuristic_no_constraint}, here we have to vary not only the maximum number $k$ of lines that can be attacked, but also the limit $Q$ on the total load of the attacked lines. In particular, we would expect the performance of the algorithms to depend heavily on $Q$. To this end, we find it meaningful to let $Q$ vary with $k$ and to set it in reference to the {\em mean} total load of $k$ randomly selected lines; i.e., we set \vspace{-1mm}
\begin{equation}
Q=Q(k)=c*k*\mathbb{E}[L]
\label{eq:limit_Q}
\vspace{-1mm}
\end{equation}
for some constant $c>0$. This choice enables us to tune $c$ to different levels and check performance in cases where (i) the total load is extremely limited (i.e., $c \ll 1$); ii) total load limit is such that heuristics that do not take load into account (such as max-$S$ or random attacks) will likely be able to select (close to) $k$ lines with total load very close to $Q$ (i.e., $c \approx 1$); or iii) total load limit is not stringent at all (i.e., $c \gg 1$) and the problem is similar to the unconstrained load case.

In the first set of experiments, 
we set $N=5000$ and generate load-free space values independently from the distributions $L\sim U[10,30]$ and $S\sim U[10,60]$. 
For brevity we consider four  values of  $c$ given at (\ref{eq:limit_Q}): $c=0.25$, $c=0.75$, $c=1.0$, $c=1.25$. The results are presented in 
Figure 
\ref{fig:last_indep} from which a number of interesting observations can be made. When $c=0.25$, i.e., when total load is extremely constrained, we see that all heuristics without a switch perform poorly and are not able to fail the whole system even until $k=3000$. This can be attributed to their inability to  attack the maximum allowed number $k$ of lines as they quickly exhaust the total load limit. On the other hand, we see that both max-$L*S$-with-switch and max-$S/L$-with-switch attacks perform much better, and despite the stringent limit on the total load are able to fail the system by attacking about 50\% more lines than required in the case where the total load is unlimited. 
When $c$ is increased to $0.75$, we see that the performance of the benchmarks improve but still  are significantly worse than the two heuristics that use the switch; in this case we also see that the max-$S/L$-with-switch attack slightly outperforms max-$L*S$-with-switch. 

With $c=1$, we see that algorithms that ignore the loads such as max-$S$ and random attacks perform as well as they do in the unconstrained case; this is expected by virtue of the law of large numbers. In particular, when $c=1$, we would expect max-$S$ to perform well since it picks the most robust lines in the system and is likely to reach the limits $k$ and $Q$ simultaneously given that $L$ is independent from $S$. Figure 
\ref{fig:last_indep_3} confirms this intuition where we also see that both heuristics with switch match the performance of the max-$S$ attack. Finally, with $c=1.25$, we get close to the unconstrained load case, and as expected see that the performance of the max-$L*S$ algorithm becomes the best. What is interesting here is that the max-$L*S$-attack-with-switch is able to match this performance, showing its versatile performance across very different cases considered here.
Overall, these experiments demonstrate that incorporating the {\em switch} significantly improves the performance and the max-$L*S$-attack performs well across different ranges of the total load limit.

As in Section~\ref{heuristic_no_constraint}, it is of interest to check the performance of these algorithms in difficult cases where load and free-space values are sorted in reverse orders. To this end, we consider one of the settings used in Figure \ref{fig:reverse_order}, and generate load and free-space independently from $L\sim U[0.4,100]$ and
  $S\sim U[0.05,150]$, and then sort them in reverse orders so that the line with maximum load gets the smallest free-space and so on and so forth. In this setting, max-$S$ and max-$S/L$ heuristics become equivalent. Also, since large $S$ values are around 150 while $L$ is limited to $100$, the lines with maximum-$C$ will be those with large $S$ (and small $L$ due to reverse ordering). As seen in Figure \ref{fig:last_inverse}, this leads to three benchmarks (max-$S,C,S/L$) performing almost equally in this setting. 
  
We see from Figure \ref{fig:last_inverse} that the max-$L*S$-attack-with-switch once again performs well. It leads to the best performance among all heuristics considered for $c=0.25$, $c=0.75$, and $c=1.0$ (equal with max-$L*S$), while coming second for $c=1.25$ after max-$L*S$; this is expected since without a stringent limit on total load, the problem gets closer to the unconstrained case where a switch is not needed.  Also, in this case we see that the performance of the max-$S$ attack (along with  max-$S/L$ and max-$C$ attacks) is rather unaffected by the load constraint. We attribute this to the fact that since load and free-space are reverse ordered, targeting max-$S$ lines is equal to targeting min-$L$ lines and even when $c=0.25$ the total load limit is not likely to be exhausted easily; i.e., the algorithm is able to choose $k$ lines without exceeding the total load limit. We think this is also the reason why the {\em switch} is not helping (and, actually hurting) the max-$S/L$ algorithm in this case.

%


\section{Simulations with UK National Grid Data}
\label{sec:UKdata}

In this section, we provide simulation results illustrating
how the attack strategies covered here performs when the load
and free space distribution are based on real data. 
We have used National Grid Electricity Ten Year Statement 2016 Model of Great Britain \cite{real_data} to generate load-free space pairs. 
To be more precise,
the load
distribution is chosen from the winter peak power flow diagram presented in \cite[Appendix C]{real_data}. For the free-space distribution,
the transmission line ratings given in \cite[Appendix B]{real_data} has been used. As in the case of previous examples,
the number of lines is taken as $N=5000$.

\begin{figure}[!b]
\vspace{-1mm}
  \centering
  \includegraphics[width=.48\textwidth]{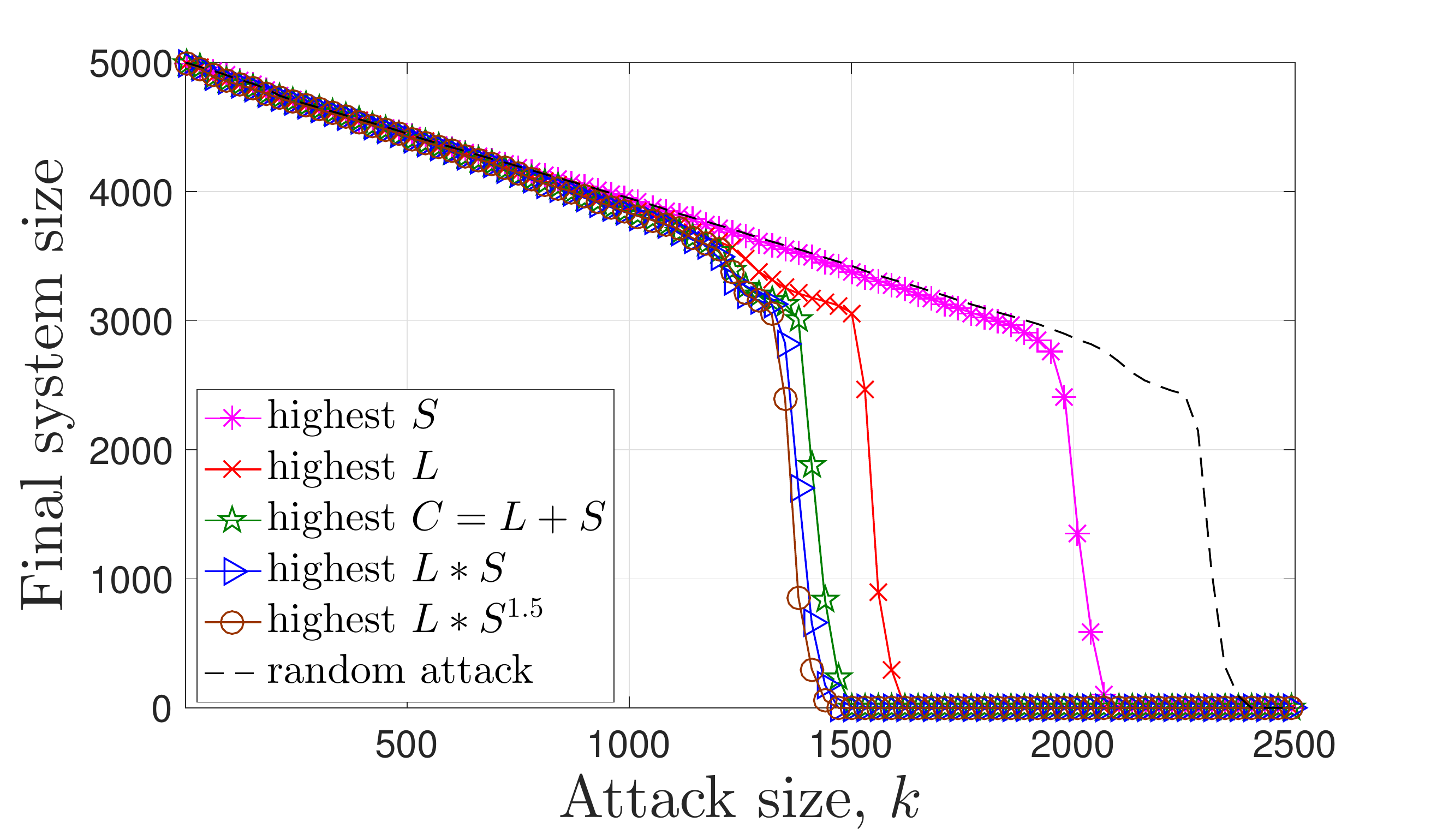}
 \vspace{-2mm} \caption{ \sl The performance comparison of different heuristic algorithms when $L,S$
follow the UK National Grid data \cite{real_data}.}
  \label{fig:UK_no_constraint}\vspace{-4mm}
\end{figure}

In Figure \ref{fig:UK_no_constraint}, we show the performance of the heuristic algorithms for the unconstrained case, i.e., for the ER-$k$ problem. The results are very similar to those obtained under synthetic load-free space distributions and demonstrate that  heuristics developed here perform well under real-world distributions as well. In particular, we see that the proposed max-$L*S$ heuristic performs better than all benchmarks considered, and its performance can further be improved by the max-$L*S^{\beta}$ attack. For the UK National Grid data, our results indicated that the best performance is obtained when $\beta = 1.5$; see also the last row in Table I.

\begin{figure*}[t] 
\vspace{-3mm}
    \centering
\subfigure[]
    {
    \includegraphics[width=0.46\linewidth]{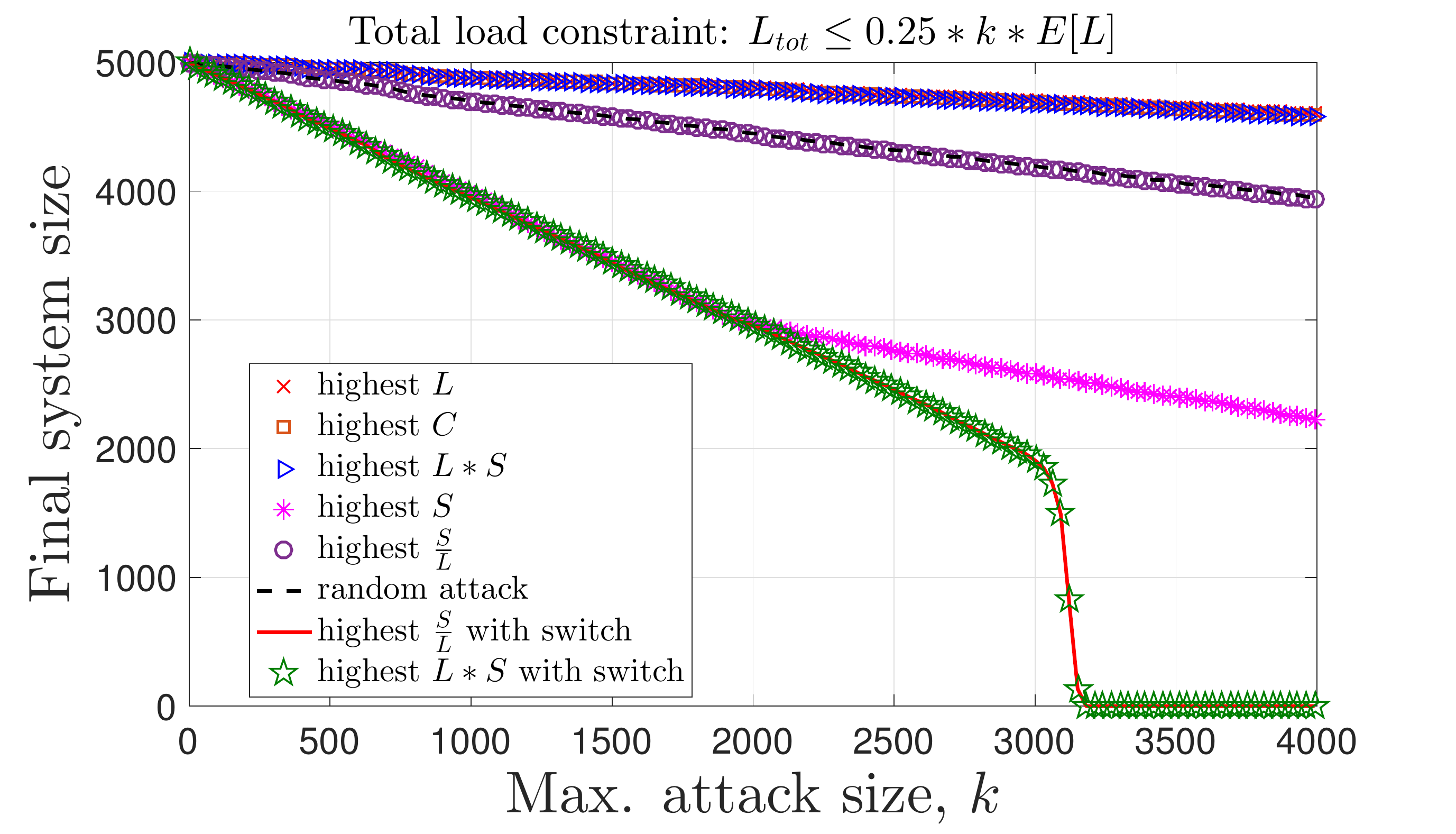} 
    \label{fig:data_1}} 
\subfigure[]
    {
    \includegraphics[width=0.46\linewidth]{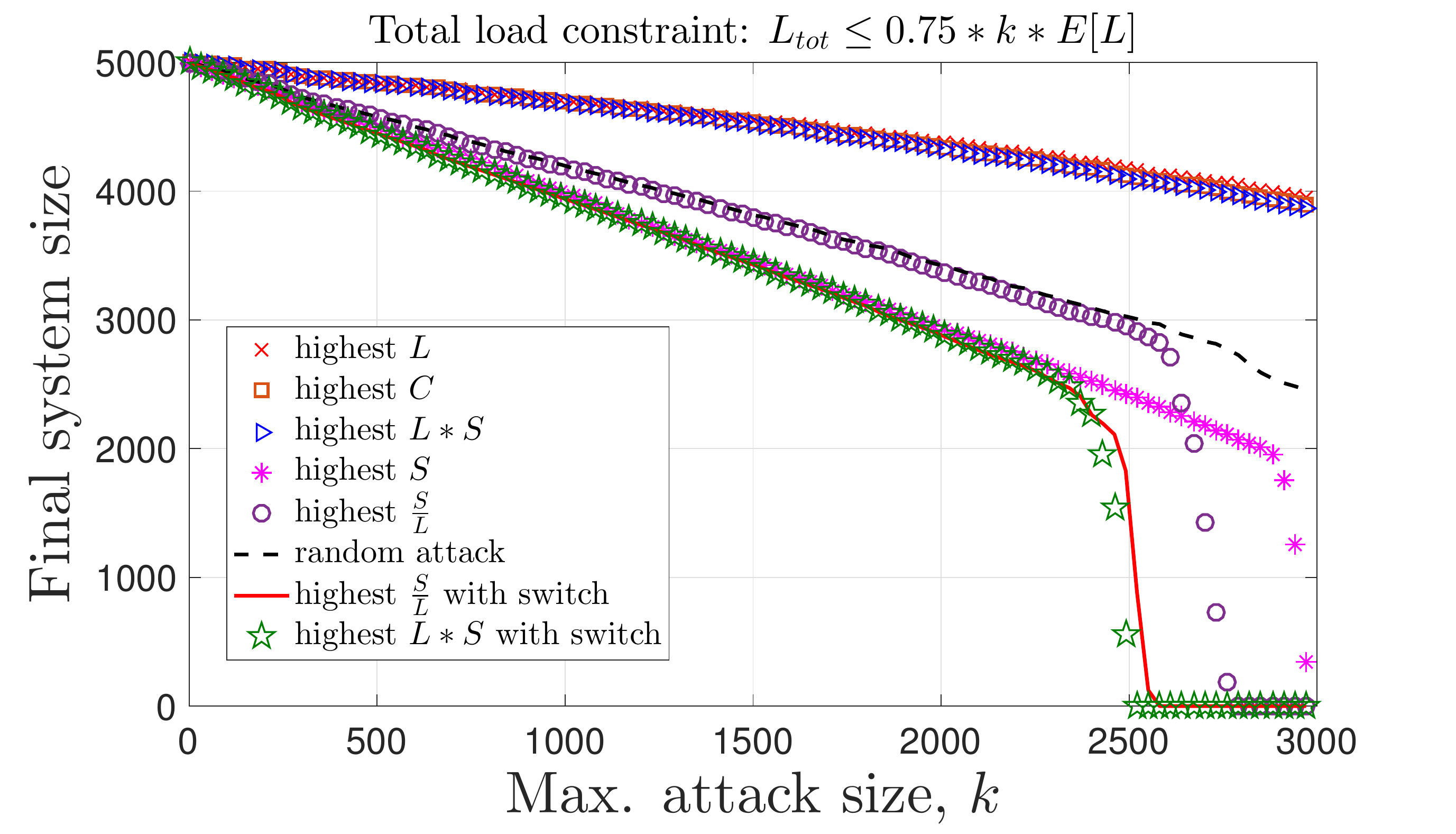} 
    \label{fig:data_2}} 
    \subfigure[]
    {
    \includegraphics[width=0.46\linewidth]{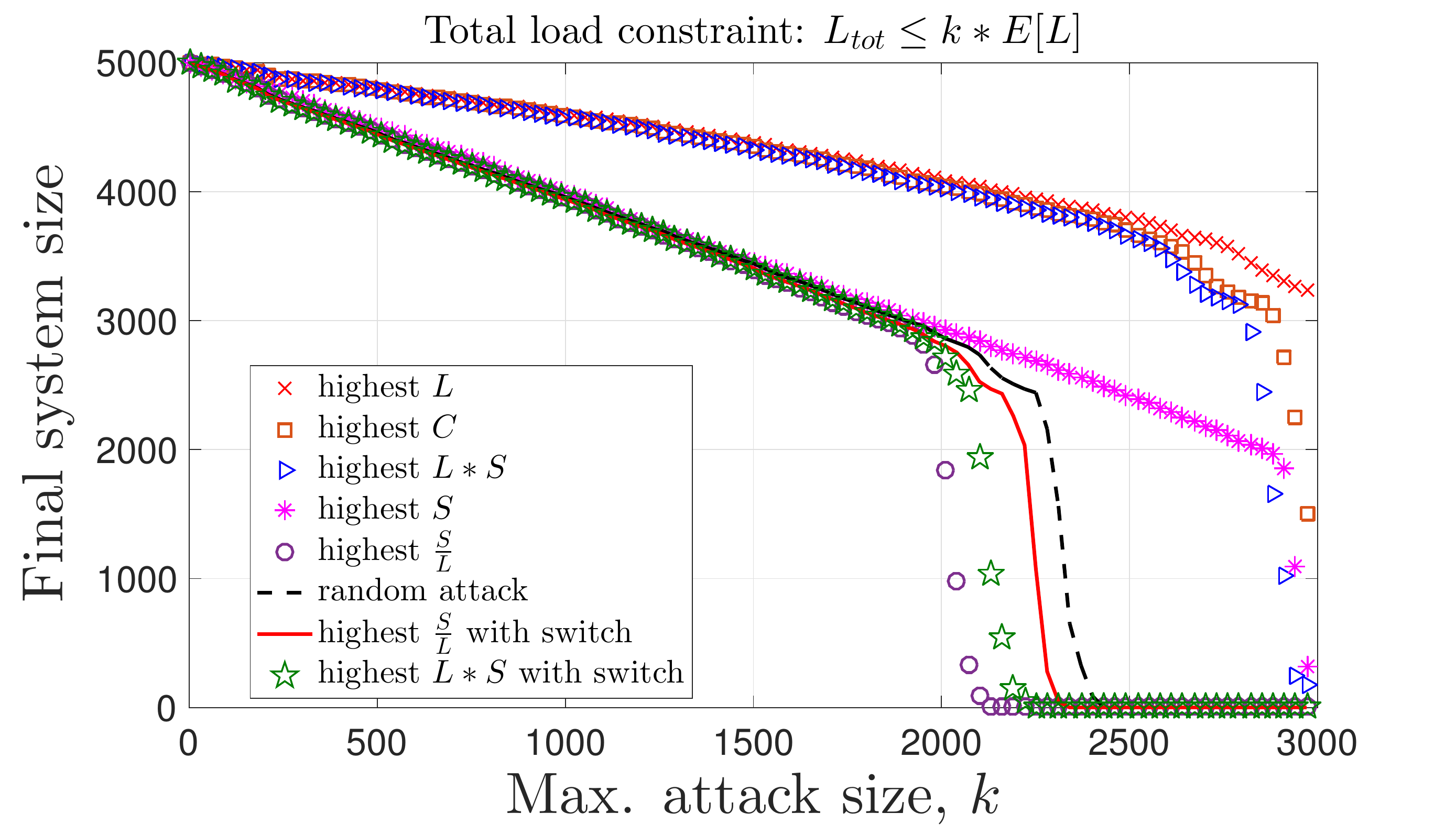} 
    \label{fig:data_3}} 
    \subfigure[]
    {
    \includegraphics[width=0.46\linewidth]{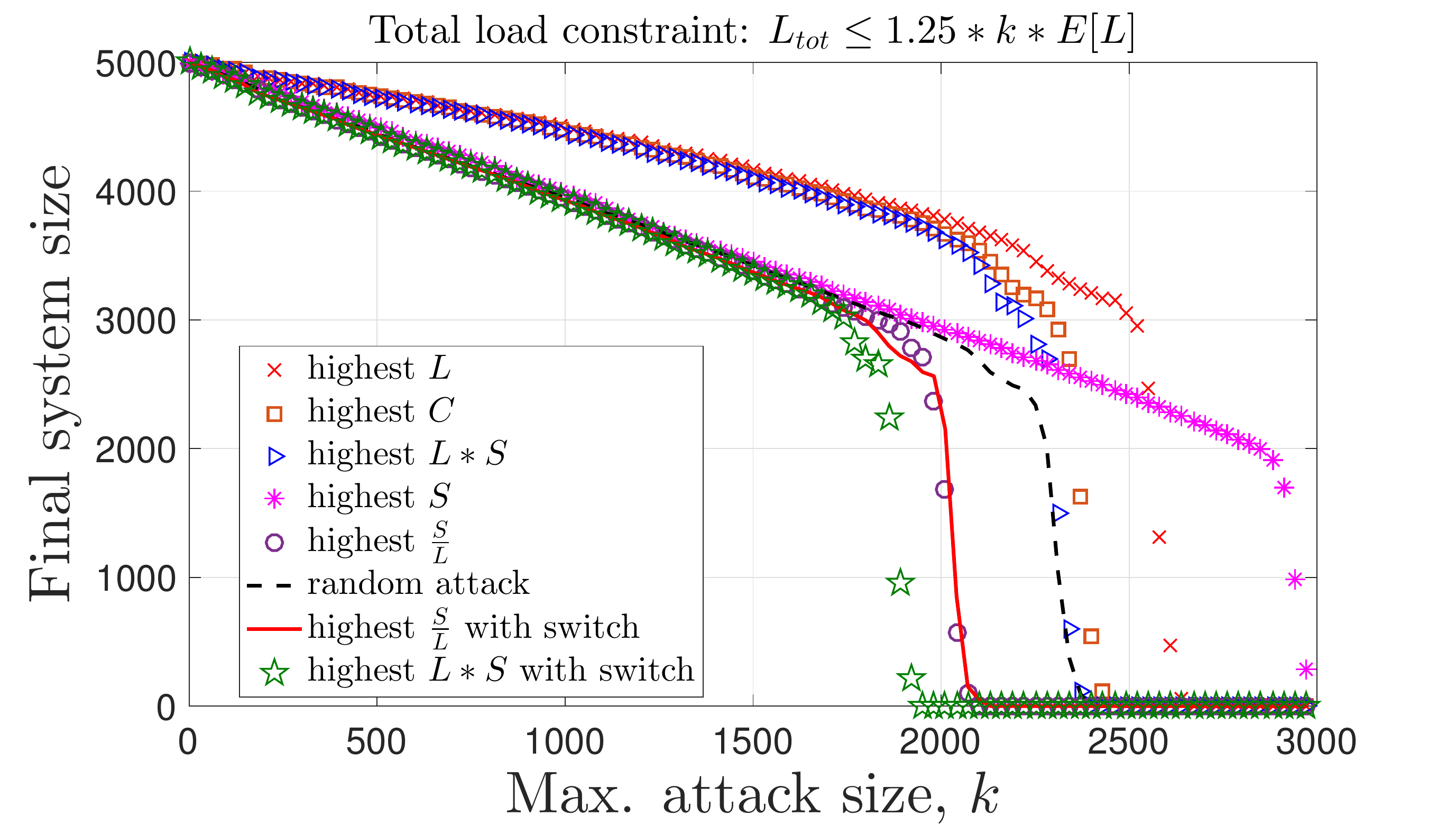} 
    \label{fig:data_4}} 
    \hspace{8mm}
\vspace{-4mm} \caption{ \sl The performance comparison of different heuristic algorithms when $L,S$ pairs are distributed according to the UK National Grid data \cite{real_data}. The attack is constrained to $k$ lines such that their total load satisfies a) $L_{\textrm{tot}} \leq 0.25*k*\mathbb{E}[L]$; b) $L_{\textrm{tot}} \leq 0.75*k*\mathbb{E}[L]$; c) $L_{\textrm{tot}} \leq 1.0*k*\mathbb{E}[L]$; d) $L_{\textrm{tot}} \leq 1.25*k*\mathbb{E}[L]$. 
}
  \label{fig:UKdata} \vspace{-4mm}
\end{figure*}

Next, we consider the ER-$k$-$k'$-$Q$ problem where 
 the total load of the attack set is bounded by $Q$.  As in Figures \ref{fig:last_indep} and \ref{fig:last_inverse},
 we set $Q$ according to (\ref{eq:limit_Q})  for several $c$ values, and compare the performance of our attack strategies with benchmarks; this time the load-free space distribution is set according to the aforementioned UK National Grid data. The results
 are depicted in Figure \ref{fig:UKdata}. Once again we see that the 
 max-$L*S$ attack {\em with switch} leads to the best overall performance among all heuristics considered; it leads to the best performance when $c=0.25$ (tied with max-$S/L$ with switch), $c=0.75$, and $c=1.25$, while coming as second for $c=1$ after max-$S/L$.  
This shows that the heuristic attacks proposed here have versatile performance also under distributions observed in real-world power systems.

\section{Conclusion}
In this paper, we  consider a flow-redistribution based model 
to understand cascading failures in load-carrying networks, e.g., electrical power systems, transportation systems, etc. 
We focus on assessing the vulnerability of such systems against adversarial attacks. 
In particular, in a system with $N$ lines with initial loads $L_1, \ldots, L_N$ and capacities $C_1,\ldots, C_N$, we study the constrained optimization problem of finding $k$ initial lines to be attacked to minimize the final number of alive lines in the system. We give optimal greedy algorithms in several special cases, and prove that a variation of the problem (with a bound on the total load of the initial attack set) is NP-Hard. Several heuristics are developed 
and their performance is compared with benchmark attacks under various settings. Overall, it is seen that the system is most vulnerable against 
attacks that target lines with maximum load free-space product $L*(C-L)$. 

There are many interesting directions to consider for future work. First, the complexity of the optimal $k$-attack problem (without a bound on the total load) is not known.  Also, with the results of this paper revealing {\em good} attack strategies, one might now seek optimal design strategies  (e.g., in the form of load-capacity distributions) that lead to maximum robustness against such attacks. It would also be interesting to 
consider new cascading failure models for flow networks that combine local and global redistribution approaches; e.g., a portion
of the failed load is redistributed in the local neighborhood according to network topology while the rest is redistributed globally. 
For this last problem, we provide preliminary simulation results in Supplementary Material.
They suggest that the mean-field (i.e., equal) redistribution model captures the qualitative behavior of system robustness well.
Finally, it might be interesting to study {\em information cascades} in social networks \cite{WattsExternal,YaganPRE,YaganCISS2012,YaganQianInfoPropLong} using the models considered here; the optimal attack problem studied here will then amount to  {\em influence maximization} problem \cite{kempe2003maximizing}.






%
\bibliographystyle{abbrv}
\bibliography{references}

\end{document}

%% file: pream.tex
\newtheorem{prop}{Proposition}[section]

\newtheorem{cor}{Corollary}

\newtheorem{lm}{Lemma}

\newtheorem{thm}{Theorem}

\newcommand{\bthm}{\begin{thm}}
\newcommand{\ethm}{\end{thm}}

\newcommand{\bcor}{\begin{cor}}
\newcommand{\ecor}{\end{cor}}
\newcommand{\bprop}{\begin{prop}}
\newcommand{\eprop}{\end{prop}}
\newcommand{\blm}{\begin{lm}}
\newcommand{\elm}{\end{lm}}
\newcommand{\beq}{\begin{equation}}
\newcommand{\eeq}{\end{equation}}
\newcommand{\ber}{\begin{eqnarray}}
\newcommand{\eer}{\end{eqnarray}}

\newenvironment{proof1}{\begin{trivlist}\item[]{\bf Proof:\hspace{2mm}}}{\hfill$\blackbox$\end{trivlist}}

\newcommand{\fsquare}{\vrule height6pt width7pt depth1pt}   

\newcommand{\myproof}{{\hfill \\ \bf Proof. \ }}           
\newcommand{\myendpf}{\hfill\fsquare \\[0.1in]}             

\newcommand{\blackbox}{\vrule height7pt width5pt depth1pt}

\newcommand{\bit}{\begin{itemize}}
\newcommand{\eit}{\end{itemize}}
\newcommand{\ben}{\begin{enumerate}}
\newcommand{\een}{\end{enumerate}}
\newcommand{\bdesc}{\begin{description}}
\newcommand{\edesc}{\end{description}}
\newcommand{\beqarrn}{\begin{eqnarray*}}
\newcommand{\eeqarrn}{\end{eqnarray*}}
\newenvironment{proofof}[1]{\begin{trivlist}\item[]{\bf Proof of #1:\hspace{2mm}
}}{\hfill\blackbox\end{trivlist}}
\newcommand{\bproofof}{\begin{proofof}}
\newcommand{\eproofof}{\end{proofof}}
\newenvironment{rem}{\begin{trivlist}\item[]{\bf
Remark:}\hspace{4mm}}{\end{trivlist}}
\newcommand{\brem}{\begin{rem}}
\newcommand{\erem}{\end{rem}}
\newenvironment{rems}{\begin{trivlist}\item[]{\bf
Remarks}\begin{itemize}}{\end{itemize}\end{trivlist}}
\newcommand{\brems}{\begin{rems}}
\newcommand{\erems}{\end{rems}}
\newtheorem{fact}{Fact}
\newcommand{\bfact}{\begin{fact}}
\newcommand{\efact}{\end{fact}}
\newtheorem{examp}{Example}
\newcommand{\bexamp}{\begin{examp}\rm}
\newcommand{\eexamp}{\end{examp}}
\newtheorem{defn}{Definition}
\newcommand{\bdefn}{\begin{defn}\rm}
\newcommand{\edefn}{\end{defn}}

\newtheorem{prob}{Problem}
\newcommand{\bprob}{\begin{prob}}
\newcommand{\eprob}{\end{prob}}

\newcommand{\bvtm}{\begin{verbatim}}
\newcommand{\bfig}{\begin{figure}}
\newcommand{\efig}{\end{figure}}
\newcommand{\bcen}{\begin{center}}
\newcommand{\ecen}{\end{center}}

\long\def\comment#1{}

%% file: math_def.tex
\def \n2{{N_0 \over 2}}

\newcommand{\bP}[1]{{\mathbb{P}}\left[{#1}\right]}
\newcommand{\bE}[1]{{\mathbb{E}}\left[{#1}\right]}

\newcommand{\1}[1]{{\bf 1}\left[#1\right]}

\def \h5{\hspace{0.5in}}